\documentclass[lettersize,journal]{IEEEtran}
\usepackage{amsmath,amsfonts}
\usepackage{algorithmic}
\usepackage{algorithm}
\usepackage{array}
\usepackage[caption=false,font=normalsize,labelfont=sf,textfont=sf]{subfig}
\usepackage{textcomp}
\usepackage{stfloats}
\usepackage{url}
\usepackage{verbatim}
\usepackage{graphicx}
\usepackage{cite}
\usepackage{cite}
\usepackage{acronym}
\usepackage{booktabs}
\usepackage{ifpdf}

\ifpdf
\else
\fi

\usepackage{cite}
\usepackage{balance}
\usepackage{bm,comment,color}

\usepackage{amsmath}

\usepackage{array}
\usepackage{amsfonts} 
\usepackage{amssymb}
\usepackage{esint} 
\usepackage{units}
\usepackage{multirow}
\usepackage{comment}

\usepackage{color}

\definecolor{purple}{RGB}{115, 0, 255}

\usepackage{pgfplots}
\usepackage{tikz}
\usetikzlibrary{calc}
\makeatletter
\newcommand{\gettikzxy}[3]{%
  \tikz@scan@one@point\pgfutil@firstofone#1\relax
  \edef#2{\the\pgf@x}%
  \edef#3{\the\pgf@y}%
}
\usetikzlibrary{spy,backgrounds}
\pgfplotsset{compat=newest}
\usetikzlibrary{plotmarks}
\usetikzlibrary{arrows.meta}
\usepgfplotslibrary{patchplots}
\usepackage{grffile}
\newlength\fheight 
\newlength\fwidth 
\usepgfplotslibrary{fillbetween}

\usepackage[a4paper, top=0.38in, bottom=1in, left=0.62in, right=0.62in, includeheadfoot]{geometry}

\allowdisplaybreaks

\acrodef{ad}[AD]{autonomous drive}
\acrodef{adas}[ADAS]{advanced driver assistance system}
\acrodef{aoa}[AOA]{angles-of-arrival}
\acrodef{aod}[AOD]{angles-of-departure}\acrodef{bs}[BS]{base station}
\acrodef{cdf}[CDF]{cumulative distribution function}
\acrodef{crb}[CRB]{Cram\'er-Rao bound}
\acrodef{dbscan}[DBSCAN]{density-based spatial clustering of applications with noise}
\acrodef{gnss}[GNSS]{global navigation satellite system}
\acrodef{gps}[GPS]{global positioning system}
\acrodef{imu}[IMU]{inertial measurement unit}
\acrodef{isac}[ISAC]{integrated sensing and communication}
\acrodef{ip}[IP]{incidence point}
\acrodef{las}[L\&S]{localization and sensing}
\acrodef{los}[LOS]{line-of-sight}
\acrodef{mae}[MAE]{mean absolute value}
\acrodef{map}[MAP]{maximum a posteriori}
\acrodef{mcrb}[MCRB]{misspecified Cram\'er-Rao bound}
\acrodef{mimo}[MIMO]{multiple-input-multiple-output}
\acrodef{mle}[MLE]{maximum likelihood estimator}
\acrodef{mlb}[MLB]{mismatched lower bound}
\acrodef{mpc}[MPC]{multipath component}
\acrodef{nlos}[NLOS]{non-line-of-sight}
\acrodef{ofdm}[OFDM]{orthogonal frequency division multiplexing}
\acrodef{prs}[PRS]{positioning reference signal}
\acrodef{pss}[PSS]{primary synchronization signal}
\acrodef{rf}[RF]{radio frequency}
\acrodef{ris}[RIS]{reconfigurable intelligent surface}
\acrodef{rss}[RSS]{received signal strength}
\acrodef{rtk}[RTK]{real-time kinematic}
\acrodef{rtt}[RTT]{round-trip-time}
\acrodef{siso}[SISO]{single-input-single-output}
\acrodef{snr}[SNR]{signal-to-noise ratio}
\acrodef{slam}[SLAM]{simultaneous localization and mapping}
\acrodef{ssb}[SSB]{synchronization signal/physical broadcast channel block}
\acrodef{tdoa}[TDOA]{time-difference-of-arrival}
\acrodef{toa}[TOA]{time-of-arrival}
\acrodef{ue}[UE]{user equipment}
\acrodef{ura}[URA]{uniform rectangular array}
\acrodef{va}[VA]{virtual anchor}

\long\def\comment#1{}

\newfont{\bbb}{msbm10 scaled 700}

\newfont{\bb}{msbm10 scaled 1100}

\newcommand{\bv}{{\bf b}}

\newcommand{\ov}{{\bf o}}
\newcommand{\pv}{{\bf p}}

\newcommand{\tv}{{\bf t}}

\newcommand{\Qm}{{\bf Q}}

\newcommand{\varphiv}{\hbox{\boldmath$\varphi$}}

\renewcommand{\arg}{{\hbox{arg}}}

\newcommand{\herm}{{\sf H}}

\begin{document}

\title{RIS-Aided Positioning Under Adverse Conditions: Interference from Unauthorized RIS}

\author{
Mengting~Li,~\IEEEmembership{Member,~IEEE},
Hui~Chen,~\IEEEmembership{Member,~IEEE}, 
Alireza~Pourafzal,~\IEEEmembership{Member,~IEEE}, Henk~Wymeersch,~\IEEEmembership{Fellow,~IEEE}
\thanks{M.~Li, H.~Chen, A. Pourafzal, and Henk are with the Department of Electrical Engineering, Chalmers University of Technology, 412 58 Gothenburg, Sweden (Email: {limeng@chalmers.se, hui.chen, alireza.pourafzal, henkw}@chalmers.se). M.~Li is also with Aalborg University, Denmark. (Email: mengli@es.aau.dk).}
\thanks{This work was supported, in part by the research grant (VIL59841) from VILLUM FONDEN, the Swedish Research Council (VR grant 2023-03821), and the SNS JU project 6G-DISAC under the EU’s Horizon Europe Research and Innovation Program under Grant Agreement No 101139130.}
}




\maketitle

\begin{abstract}

Positioning technology, which aims to determine the geometric information of a device in a global coordinate, is a key component in integrated sensing and communication systems.
In addition to traditional active anchor-based positioning systems, reconfigurable intelligent surfaces (RIS) have shown great potential for enhancing system performance. 
However, their ability to manipulate electromagnetic waves and ease of deployment pose potential risks, as unauthorized RIS may be intentionally introduced to jeopardize the positioning service. Such an unauthorized RIS can cause unexpected interference in the original localization system, distorting the transmitted signals, and leading to degraded positioning accuracy.
In this work, we investigate the scenario of RIS-aided positioning in the presence of interference from an unauthorized RIS. Theoretical lower bounds are employed to analyze the impact of unauthorized RIS on channel parameter estimation and positioning accuracy. Several codebook design strategies for unauthorized RIS are evaluated, and various system arrangements are discussed. The simulation results show that an unauthorized RIS path with a high channel gain or a delay similar to that of legitimate RIS paths leads to poor positioning performance. Furthermore, unauthorized RIS generates more effective interference when using directional beamforming codebooks compared to random codebooks.
\end{abstract}


\section{Introduction}

\Ac{isac} is envisioned as a core function in modern communication systems, enabling the dual capability of wireless networks to transmit data while simultaneously sensing the environment~\cite{liu2022integrated}. Positioning, a process of estimating the location (and possibly orientation and velocity) of a device from radio measurements, plays a pivotal role in ISAC, assisting communication systems in radio
resource management, beamforming, mobility management, etc~\cite{behravan2022positioning}. By utilizing uplink or downlink pilot signals, geometrical relationships, such as angle and time between a connected \ac{ue} and \ac{bs} anchors can be inferred for a further position estimate~\cite{dwivedi2021positioning}. In practical scenarios with multipath, positioning often works in tandem with mapping, unlocking new applications such as digital twins, autonomous systems, and collaborative robots, which rely on accurate information on positioning and the surrounding environment~\cite{gao2024localization,ge2024batch}.

The reconfigurable intelligent surface (RIS) has emerged as a transformative technology in wireless communication systems, with the ability to reshape channels using its configurable elements~\cite{pan2022overview,lian2024physics}. 
RIS can enhance positioning accuracy by working as additional passive anchors and providing high-resolution angular information, leveraging its large number of RIS elements~\cite{chen2024multi}. Additionally, RIS can be seamlessly integrated into a wide range of localization systems, from SISO to MIMO configurations, expanding capabilities from position estimation to orientation estimation~\cite{keykhosravi2023leveraging}. The deployment of multiple RIS can further improve coverage while reducing network costs,  at the expense of overhead~\cite{ma2024multi}. In out-of-coverage scenarios, sidelink communication between UEs can leverage RIS anchors to maintain localization services~\cite{chen2024multi}. Furthermore, advanced RIS implementations, such as near-field processing~\cite{chen20246g}, integration into nonterrestrial networks~\cite{zheng2024leo}, and innovative RIS types (e.g., active RIS~\cite{qiu2024active} and beyond diagonal RIS~\cite{raeisi2024efficient}), further amplify the potential of ISAC systems to operate in complex environments.

Despite the importance of location information, the misuse of inaccurate location information can have catastrophic consequences, especially for safety-critical applications, such as autonomous vehicles and industrial automation~\cite{whiton2022cellular}. From the system level, communication resources for pilot signals, such as bandwidth and transmit power, can directly affect localization performance.  However, these resource constraints should be considered during the system design phase, and are usually flexible to control during localization operations~\cite{dai2014distributed}. Another source of errors lies in the prior knowledge of the anchors, which consists of two major factors: geometry error and hardware impairment~\cite{ghazalian2024calibration}. 
A certain level of geometry errors (e.g., a few degrees of orientation or meters of position) may have a limited impact on communications, where end-to-end channel matters rather than the position information embedded in the channel. In contrast, these errors directly affect the global coordinate of the target~\cite{ge2024experimental}.
Regarding hardware impairments that exist in communication systems, certain types of factors, such as mutual coupling and antenna displacement error, have shown a greater impact in localization and require careful calibration~\cite{chen2023modeling}.

Beyond system-introduced errors, adversarial threats, such as intentional jamming and spoofing, also pose significant risks to localization services~\cite{goztepe2021localization}. These malicious attacks range from random Gaussian signals~\cite{gezici2016jamming} to dedicated designed pilots~\cite{zhang2024privacy}, affecting both model-based algorithms and learning-based localization systems~\cite{huang2024attacking}.
Without a proper response to such attacks, the availability of the positioning service can be jeopardized, and unsafe decision making based on unaware and erroneous estimations can occur.
With the help of secure sensors, attack detectors are proposed to detect malicious sensors by scrutinizing the existence of geometric inconsistencies~\cite{zhang2018attack}. Similarly, in the measurement consistency check, information-theoretic tools can be leveraged to detect attacks and support navigation under adverse conditions~\cite{michieletto2022robust}. However, such solutions often require additional devices and may not be practical in reality, highlighting the need for efficient and scalable approaches.

In RIS-aided systems, initial work has investigated the geometry error of RIS anchors in positioning by implementing \ac{mcrb} for mismatch analysis~\cite{zheng2023misspecified}. Joint \ac{ue} localization and RIS calibration algorithms have been proposed to tackle this challenge~\cite{zheng2023jrcup}.
Regarding the hardware imperfection of the RIS, a pilot-based phase calibration task is addressed by alternating optimization~\cite{zhang2022phase}. In~\cite{ozturk2024ris}, joint localization and failure diagnosis strategies are developed to detect failing pixels while simultaneously locating the UE with high accuracy. However, existing works only consider a limited number of impairment types, requiring practical and scalable calibration algorithms that can deal with a wide range of scenarios.

Treating an RIS as an adversary in communication systems is not new, which requires specific security solutions to address the vulnerability of
physical layer security attacks~\cite{strinati2021reconfigurable}. In ~\cite{alakoca2022metasurface}, several metasurface manipulation attacks have been identified and classified, showing that the bit error rate (BER) of the legitimate user pair is extremely reduced.
The impact of RIS on the \ac{isac} multi-user MIMO system has been studied with an emphasis on \ac{snr}~\cite{rivetti2024destructive}. Although \ac{snr} directly affects the probability of detection and the position estimate, the quantitative impact on position performance cannot be achieved.
Specifically, unauthorized RIS deployments can distort pilot signals from legitimate RIS anchors, directly affecting channel parameter extraction tasks and hence the accuracy and reliability of localization estimations. Unfortunately, the impact, interference strategies, and countermeasures in RIS-aided localization systems have not been studied in the existing literature.

In this work, we will evaluate positioning performance under adverse conditions, where an unauthorized RIS interferes with a legitimate RIS-aided localization system.
The contributions of this work can be summarized as follows:
\begin{itemize}
    \item \textbf{Formulating a scenario of RIS-aided positioning under adverse conditions with several attack strategies proposed.} In this case, the legitimate RIS acts as a passive anchor in the original localization system, facilitating the estimation of the UE position and clock offset. In the meantime, an unauthorized RIS is controlled by a malicious agent, aiming to degrade system performance without being noticed.   
    \item \textbf{Performing theoretical analysis on systems with and without interferences.} 
    The CRB is derived based on the mismatched / assumed model, that is, an interference-free localization system. In addition, MCRB-based theoretical analyses are performed to assess channel parameter estimation and positioning accuracy in the presence of unknown interference. The derived bounds can help quantify the performance degradation caused by the unauthorized RIS and serve as a benchmark for evaluating localization algorithms.   
    \item \textbf{Developing a robust localization algorithm.} We develop a low-complexity algorithm to enhance channel parameter estimation accuracy under adverse interference. Furthermore, the \ac{mle} is employed to obtain optimized estimation results based on coarse estimates from the low complexity estimation process.
    \item \textbf{Evaluation and validation of the theoretical analysis and developed algorithms:} Extensive simulations are conducted for multiple purposes:
a) Demonstrating the effectiveness of the derived bounds and the developed localization algorithms.
b) Investigating the impact of the beamforming codebooks used by unauthorized RIS.
c) Evaluating various deployment scenarios (e.g., different placement of legitimate and unauthorized RISs).
\end{itemize}


\textit{Notations:} Vectors are represented by bold lowercase letters, matrices by bold uppercase letters, and scalars by italicized letters. Unless stated otherwise, all vectors are column vectors. The transpose, inverse, Hermitian, trace, and complex conjugate operations are denoted as \( (\cdot)^{\top} \), \( (\cdot)^{-1} \), \( (\cdot)^{\mathsf{H}} \), \( \text{tr}(\cdot) \), and \( (\cdot)^{\ast} \), respectively. The Hadamard product is denoted by \( \odot \). The notation \( [\cdot]_i \) refers to the \(i\)-th element of a vector, while \( [\cdot]_{i,j} \) represents the element at the \(i\)-th row and \(j\)-th column of a matrix. Additionally, the subscript \( i:j \) denotes all elements between indices \( i \) and \( j \).

    
\section{System Model and Problem Statement}

\begin{figure}
    \centering
    \includegraphics[width=0.9\linewidth]{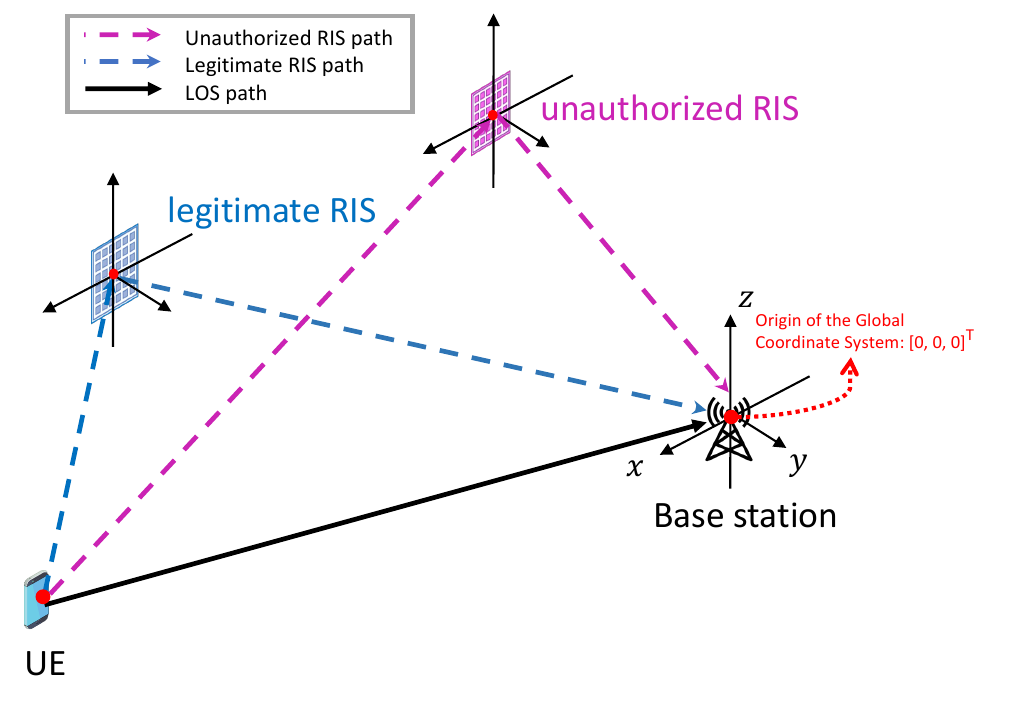}
    \caption{An illustration of a RIS-aided localization system with the presence of an unauthorized RIS. In the absence of the unauthorized RIS, positioning is achieved via signals from the LOS path and the legitimate RIS path. }
    \label{fig:system}
\end{figure}
We consider a \ac{siso} RIS-aided localization system \cite{elzanaty2021reconfigurable,keykhosravi2022ris} with the presence of an unauthorized RIS in a 3D scenario, as shown in Fig. \ref{fig:system}. The purpose of the legitimate RIS is to assist in estimating the position of the UE, while the unauthorized RIS remains undetected by the localization system and seeks to introduce unfavorable interference, degrading localization performance.
\subsection{Geometry and Signal Model}
The single-antenna BS and UE are located at $\pv_{\rm B}$, $\pv_{\rm U} \in \mathbb{R}^3$, respectively. The legitimate RIS contains $N$ passive reflective elements and its position and orientation are denoted by 
$\pv_{\rm R_{L}} \in \mathbb{R}^3$ and $\ov_{\rm R_{L}} \in \mathbb{R}^3$, respectively. The unauthorized RIS contains $M$ passive reflective elements and its position and orientation are denoted by $\pv_{\rm R_{U}} \in \mathbb{R}^3$ and $\ov_{\rm R_{U}} \in \mathbb{R}^3$, respectively. Note that the Euler angle vectors $\ov_{\rm R_{L}}$ and $\ov_{\rm R_{U}}$ are used to determine the rotation matrices $\Qm_{\rm R_{L}}$ and $\Qm_{\rm R_{U}}$. The beamforming performances of the RIS will be determined by the phase modulations applied to the RIS elements, i.e., RIS beamforming codebooks. We assume a wideband uplink\footnote{Equivalent to downlink positioning for SISO systems} system which adopts $K$ subcarriers and transmits $G$ orthogonal frequency division multiplexing (OFDM) symbols for each localization task. The received signal matrix $\mathbf{Y} \in \mathbb{C}^{K \times G}$ is
\begin{equation}\label{eq:recSig}
    \mathbf{Y}  = \mathbf{Y_{\rm{u}}} + \mathbf{Y_{\rm{R_{L}}}} + \mathbf{Y_{\rm{R_{U}}}} + \mathbf{N} 
\end{equation}
where \( \mathbf{Y_{\rm{u}}} = \mathbf{H_{\rm{u}}} \odot \mathbf{X} \), \( \mathbf{Y_{\rm{R_{L}}}} = \mathbf{H_{\rm{R_{L}}}} \odot \mathbf{X} \), and \( \mathbf{Y_{\rm{R_{U}}}} = \mathbf{H_{\rm{R_{U}}}} \odot \mathbf{X} \) represent the signals received from the UE through the line of sight (LOS) channel, the legitimate RIS channel, and the unauthorized RIS channel, respectively, and \( \mathbf{H_{\rm{u}}} \), \( \mathbf{H_{\rm{R_{L}}}} \), and \( \mathbf{H_{\rm{R_{U}}}} \) denote the corresponding channel matrices for each received signal.
The noise matrix $\mathbf{N} \in \mathbb{C}^{K \times G}$ denotes the additive white Gaussian noise with zero mean and variance $\sigma^2$. Moreover, the transmitted pilot signal matrix $\mathbf{X}\in \mathbb{C}^{K \times G}$ is defined as $ \mathbf{X} = \sqrt{P}\mathbf{x} \mathbf{1}_{G}^{\mathrm{T}}$, where $\mathbf{x} \in \mathbb{C}^{K \times 1}$ is the transmitted symbol vector. The same symbols are assumed for different transmissions and the transmit power for each transmission is $P$. Given model~\eqref{eq:recSig}, the received signal consists of contributions from different channels. In the absence of an unauthorized RIS, the received signal matrix is interference-free and given by
\begin{equation}\label{eq:mismatched_model}
\tilde{\mathbf{Y}} = \mathbf{Y_{\rm{u}}} + \mathbf{Y_{\rm{R_{L}}}}  + \mathbf{N} \in \mathbb{C}^{K \times G}.
\end{equation}

The channel matrix which represents the LOS channel between the UE and BS is defined as
\begin{equation}\label{eq:channel_los}
  \mathbf{H_{\rm{u}}} = \alpha_u\mathbf{d}(\tau_u)\mathbf{1}_{G}^{\mathrm{T}}  ,
\end{equation}
where $\alpha_u$ and $\tau_u$ are the complex channel gain and the corresponding propagation delay, respectively. The channels for legitimate and unauthorized RIS are defined as 
\begin{equation}\label{eq:channel_RL}
  \mathbf{H_{\rm{R_{L}}}} = (\alpha_{{r_{l}}}\mathbf{d}(\tau_{r_l})\mathbf{1}_{G}^{\mathrm{T}}) \odot \mathbf{A}(\boldsymbol{\varphi}_{D}^{r_l}, \boldsymbol{\varphi}_{A}^{r_l}) ,
\end{equation}
\begin{equation}\label{eq:channel_RU}
  \mathbf{H_{\rm{R_{U}}}} = (\alpha_{{r_{u}}}\mathbf{d}(\tau_{r_u})\mathbf{1}_{G}^{\mathrm{T}}) \odot \mathbf{A}(\boldsymbol{\varphi}_{D}^{r_u}, \boldsymbol{\varphi}_{A}^{r_u}), 
\end{equation}
where the complex channel gains for legitimate RIS and unauthorized RIS paths are denoted as $\alpha_{{r_{l}}}$ and $\alpha_{{r_{u}}}$, respectively. The corresponding delays for these two RIS paths are represented as $\tau_{r_l}$ and $\tau_{r_u}$. The $k$-th element of the delay vector represents the propagation delay of a specific path (i.e., LOS path, legitimate RIS path, or unauthorized RIS path) at the $k$-th subcarrier, which can be determined by

\begin{equation}
    [\mathbf{d}(\tau)]_k = e^{-j2\pi k \Delta f \tau}.
\end{equation}
Here, $\Delta f$ is the sub-carrier spacing, and the delays of different paths are determined by the propagation distances and the clock offset $B$ (unit: m)\footnote{The clock offset $B$ is converted to meters for easier visualization of the results presented in Section \ref{sec:simulation}.} denoted as
\begin{equation}
    \tau_{u} = \frac{\| \mathbf{p}_{\rm{B}} - \mathbf{p}_{\rm{U}} \| + B}{c},
    \label{eq:tau_los}
\end{equation}
\begin{equation}
    \tau_{r_l} = \frac{\| \mathbf{p}_{\rm{B}} - \mathbf{p}_{\rm{R_L}} \| + \| \mathbf{p}_{\rm{U}} - \mathbf{p}_{\rm{R_L}} \| + B}{c},
    \label{eq:tau_ris_l}
\end{equation}
\begin{equation}
    \tau_{r_u} = \frac{\| \mathbf{p}_{\rm{B}} - \mathbf{p}_{\rm{R_U}} \| + \| \mathbf{p}_{\rm{U}} - \mathbf{p}_{\rm{R_U}} \| + B}{c}.
\end{equation}
To facilitate the discussion of the delay results in Section~\ref{sec:simulation}, we define $ d_u = \tau_{u} \cdot c $, $ d_{r_l} = \tau_{r_l} \cdot c $, and $ d_{r_u} = \tau_{r_u} \cdot c $ to represent the pseudoranges corresponding to the LOS path, the legitimate RIS-assisted path, and the unauthorized RIS-assisted path, respectively.

The matrix $\mathbf{A}(\boldsymbol{\varphi}_{D},\boldsymbol{\varphi}_{A})\in \mathbb{C}^{K \times G}$ includes both the array weights caused by the different locations of the RIS elements and the phase modulations assigned to the RIS elements. We assume RIS phase modulations are designed based on the \ac{aoa} and \ac{aod}, and each entry of this matrix can be obtained by
\begin{equation}\label{eq:array_weights}
   [\mathbf{A}(\boldsymbol{\varphi}_{D}, \boldsymbol{\varphi}_{A})]_{k,g} = \boldsymbol{\omega}_g^{\mathrm{T}}(\mathbf{a}(\boldsymbol{\varphi}_{D}) \odot \mathbf{a}(\boldsymbol{\varphi}_{A})),
\end{equation}
where $\boldsymbol{\omega}_g \in \mathbb{C}^{N \times 1}$ is the phase modulation vector with unit norm elements ($|w_{g,n}|=1$) . 
The array steering vector under the far-field condition for a certain direction is defined as
\begin{equation}\label{eq:steer_vector}
    \mathbf{a}(\boldsymbol{\varphi}) = e^{j\frac{2\pi f_c}{c}\mathbf{Z}^{\mathrm{T}}\mathbf{t}(\boldsymbol{\varphi})},
\end{equation}
where matrix $\mathbf{Z} = [\mathbf{z}_1, \ldots, \mathbf{z}_N] \in \mathbb{R}^{3 \times N}$ determines the locations of each RIS element under the local coordinate system. The direction vector $\mathbf{t}(\boldsymbol{\varphi})$, representing a specific direction in the local coordinate system, determined by an azimuth angle $\phi$ and an elevation angle $\theta$.
We use $\varphiv_{A}^{r_l}$ and $\varphiv_{A}^{r_u}$ denote the AOA of the UE to the legitimate RIS and unauthorized RIS, respectively, while $\varphiv_{D}^{r_l}$ and $\varphiv_{D}^{r_u}$ are the AOD from these RIS plates to the BS. 
The direction vectors for these paths are provided in (\ref{eq:dir_vec_rl}) and (\ref{eq:dir_vec_ru}), with similar calculations for $\mathbf{t}(\boldsymbol{{\varphi}}_{D}^{r_l})$ and $\mathbf{t}(\boldsymbol{{\varphi}}_{D}^{r_u})$.
\begin{equation}\label{eq:dir_vec_rl}
\mathbf{t}(\boldsymbol{{\varphi}}_{A}^{r_l}) =  \mathbf{Q}_{\rm{R_L}}^{-1} \frac{\mathbf{p}_{\rm{U}} - \mathbf{p}_{\rm{R_L}}}{\| \mathbf{p}_{\rm{U}} - \mathbf{p}_{\rm{R_L}} \|} 
= 
\begin{bmatrix}
\cos(\phi_{A}^{r_l}) \cos(\theta_{A}^{r_l}) \\
\sin(\phi_{A}^{r_l}) \cos(\theta_{A}^{r_l}) \\
\sin(\theta_{A}^{r_l})
\end{bmatrix},
\end{equation}
\begin{equation}\label{eq:dir_vec_ru}
\mathbf{t}(\boldsymbol{{\varphi}}_{A}^{r_u}) =  \mathbf{Q}_{\rm{R_U}}^{-1} \frac{\mathbf{p}_{\rm{U}} - \mathbf{p}_{\rm{R_U}}}{\| \mathbf{p}_{\rm{U}} - \mathbf{p}_{\rm{R_U}} \|} 
= 
\begin{bmatrix}
\cos(\phi_{A}^{r_u}) \cos(\theta_{A}^{r_u}) \\
\sin(\phi_{A}^{r_u}) \cos(\theta_{A}^{r_u}) \\
\sin(\theta_{A}^{r_u})
\end{bmatrix}.
\end{equation}

\subsection{RIS Beamforming Design}

The coefficients assigned to the RIS elements directly 
impact the accuracy of the localization systems. When both legitimate and unauthorized RIS are present in the same system, the intertwined channel makes positioning task intricate.
In the following, we discuss the reasonable assumptions for the beamforming design of both the legitimate and unauthorized RIS, as well as the interference impact on post-processing.

\subsubsection{Time-orthogonal RIS Profile}

The time-orthogonal RIS profile is a highly effective method to distinguish between the LoS path and the legitimate RIS path, making it widely used in RIS-aided positioning \cite{chen2024multi,keykhosravi2022ris}. This technique involves applying a temporally orthogonal phase profile (based on the original beamforming codebook) to the RIS, followed by a post-processing step at the receiver. In our scenario, it is reasonable to assume that the legitmate RIS adopts the time-orthogonal profile, as its primary purpose is to assist in positioning. 
In the following, we present the principles of the time-orthogonal RIS profile used in our localization system. 

For ease of notation, we assume $G$ is an even number and define $ \mathbf{B}^{r_l} = [\bv_1^{r_l}, \ldots, \bv_{\frac{G}{2}}^{r_l}] \in \mathbb{C}^{N\times G/2} $ as the beamforming codebook consisting of a set of codewords used by the legitimate RIS. The extension to odd $ G $ follows similarly with appropriate modifications.
The phase modulation vector $\boldsymbol{\omega}_g^{r_l}$ of the legitimate RIS using a time-orthogonal profile is set as
\begin{equation}
\boldsymbol{\omega}_{1:\frac{G}{2}}^{r_l} = \bv_{1:\frac{G}{2}}^{r_l},
\boldsymbol{\omega}_{(\frac{G}{2}+1):G}^{r_l} = -\bv_{1:\frac{G}{2}}^{r_l},
\end{equation}
From (\ref{eq:channel_los}) and (\ref{eq:channel_RL}), we can obtain
\begin{equation}
 [\mathbf{Y_{\rm{u}}}]_{:,k+\frac{G}{2}}=[\mathbf{Y_{\rm{u}}}]_{:,k} ,
\end{equation}
\begin{equation}
[\mathbf{Y_{\rm{R_L}}}]_{:,k+\frac{G}{2}}=-[\mathbf{Y_{\rm{R_L}}}]_{:,k}. 
\end{equation}
If there is no unauthorized RIS in the localization system, the LOS path and the legitimate RIS path can be separated as
\begin{equation}\label{eq:los_mismatched}
 \mathring{\mathbf{Y}}_{\rm{u}} =([\mathbf{Y}]_{:,k} + [\mathbf{Y}]_{:,k+\frac{G}{2}})/2  
 = [\mathbf{Y_{\rm{u}}}]_{:,k} + \mathring{\mathbf{N}}_{\rm{u}}
\end{equation}
\begin{equation}\label{eq:ris_mismatched}
 \mathring{\mathbf{Y}}_{\rm{R_L}} =([\mathbf{Y}]_{:,k} - [\mathbf{Y}]_{:,k+\frac{G}{2}})/2  
 = [\mathbf{Y_{\rm{R_L}}}]_{:,k} + \mathring{\mathbf{N}}_{\rm{R_L}}
\end{equation}
 where $\mathring{\mathbf{Y}}_{\rm{u}}$, $ \mathring{\mathbf{Y}}_{\rm{R_L}} \in \mathbb{C}^{K \times \frac{G}{2}}$ are used as updated received signal blocks for positioning. Since the sum of independent Gaussian variables remains Gaussian, the noise matrices $\mathring{\mathbf{N}}_{\rm{u}}$ and $\mathring{\mathbf{N}}_{\rm{R_L}}$ are additive white Gaussian matrices. The employment of orthogonal RIS beamforming design distinguishes RIS path and non-RIS paths.

 \subsubsection{The Impact of Unauthorized RIS}
 When the legitimate RIS employs the time-orthogonal profile, the presence of an unauthorized RIS inevitably impacts the accuracy of path separation. The received signal blocks in \eqref{eq:los_mismatched} and \eqref{eq:ris_mismatched} are updated as
\begin{equation}\label{eq:updated_yu}
 \mathring{\mathbf{Y}}_{\rm{u}} = [\mathbf{Y_{\rm{u}}}]_{:,k} + ([\mathbf{Y_{\rm{R_U}}}]_{:,k} + [\mathbf{Y_{\rm{R_U}}}]_{:,k+\frac{G}{2}})/2 + \mathring{\mathbf{N}}_{\rm{u}}
\end{equation}
\begin{equation}\label{eq:updated_yrl}
 \mathring{\mathbf{Y}}_{\rm{R_L}} = [\mathbf{Y_{\rm{R_L}}}]_{:,k} + ([\mathbf{Y_{\rm{R_U}}}]_{:,k} - [\mathbf{Y_{\rm{R_U}}}]_{:,k+\frac{G}{2}})/2 + \mathring{\mathbf{N}}_{\rm{R_L}}.
\end{equation}

The localization system aims to separate the propagation paths through post-processing at the receiver, as shown in (\ref{eq:los_mismatched}) and (\ref{eq:ris_mismatched}). However, when accounting for interference from unauthorized RIS, the actual received signal blocks are given by (\ref{eq:updated_yu}) and (\ref{eq:updated_yrl}). It can be inferred that the selection of the beamforming codebook employed by the unauthorized RIS, i.e., $\mathbf{B}^{r_u} = [\bv_1^{r_u}, \ldots, \bv_{G}^{r_u}]\in \mathbb{C}^{N\times G}$, will have a significant impact on the effectiveness of path separation, affecting the overall positioning accuracy. 

\subsubsection{Codebook Selection Strategies}
For the original beamforming codebook of the legitimate RIS, i.e., $\mathbf{B}^{r_l}$, either a random phase codebook or a directional codebook can be used, depending on the system's knowledge of the UE. In this paper, we focus primarily on the random beamforming codebook, as localization systems typically lack prior knowledge of the UE.
The unauthorized RIS is typically controlled by malicious agents aiming to degrade the positioning performance of the original localization system. The choice of the beamforming codebook for the unauthorized RIS largely depends on the extent of knowledge the malicious agent has about the positioning system.
\paragraph{Random Codebook} The random codebook is typically selected for unauthorized RIS when the malicious agent has limited knowledge about the positions of BS and UE. In this case, the phases assigned to the elements of the unauthorized RIS are uniformly distributed random values within the range of $[0, 2\pi)$, varying between different elements and transmissions. When the legitimate localization system conducts the post-processing to separate the LOS path and legitimate RIS path, it can be observed from (\ref{eq:updated_yu}) and (\ref{eq:updated_yrl}) that there will be both residuals for LOS and RIS paths since $[\mathbf{Y_{\rm{R_U}}}]_{:,k} \neq [\mathbf{Y_{\rm{R_U}}}]_{:,k+\frac{G}{2}}$.

\paragraph{Directional codebook} In the following, three directional codebooks are discussed: the basic directional codebook (SDC), the random angle directional codebook (RADC), and the random phase directional codebook (RPDC).
\begin{itemize}
    \item 
    With the prior knowledge of the accurate position of BS and UE, a basic directional codebook can be used to maximize the power of the interference signal on the BS side. Based on (\ref{eq:array_weights}) and (\ref{eq:steer_vector}), the codeword of SDC for the $g$-th transmission is given by 
\begin{equation}\label{eq:codebook_dir0}
    \mathbf{b}_g^{r_u} =  \mathbf{a}^{*}(\boldsymbol{\varphi}) = e^{-j\frac{2\pi f_c}{c}\mathbf{Z}^{\mathrm{T}} (\mathbf{t}(\boldsymbol{{\varphi}}_{A}^{r_u} + \mathbf{t}(\boldsymbol{{\varphi}}_{D}^{r_u}))}.
\end{equation}
This directional codebook is straightforward in design. However, assuming precise knowledge of both the BS and UE positions may not always be practical. 
\item To enhance interference and adopt a more practical assumption, two improved directional codebooks are proposed. In the first modification, the AOA from the UE to the unauthorized RIS, as known by the malicious agent, spans a range of angles rather than a single exact value. Therefore, the AOA used in (\ref{eq:codebook_dir0}) should be updated as $\tilde{\boldsymbol{{\varphi}}}_{A}^{r_u} = [\phi_{A}^{r_u}+\Delta \phi_g, \theta_{A}^{r_u}+\Delta \theta_g]^{\rm{T}}$ and $\Delta \phi_g$, $\Delta \theta_g$ are random angle shifts within a certain range. 
The codeword of RADC is derived as
 \begin{equation}\label{eq:codebook_dir1}
    \mathbf{b}_g^{r_u} =  e^{-j\frac{2\pi f_c}{c}\mathbf{Z}^{\mathrm{T}} (\mathbf{t}(\tilde{\boldsymbol{{\varphi}}}_{A}^{r_u} + \mathbf{t}(\boldsymbol{{\varphi}}_{D}^{r_u}))}.
\end{equation}
\item
An alternative way to strengthen interference impacts on the legitimate RIS path is to add a random phase $\varphi_g \sim \mathcal{U}(0, 2\pi]$ to the directional codeword obtained in (\ref{eq:codebook_dir0}), i.e.,
     \begin{equation}\label{eq:codebook_dir2}
    \mathbf{b}_g^{r_u} =  e^{-j\frac{2\pi f_c}{c}\mathbf{Z}^{\mathrm{T}} (\mathbf{t}(\boldsymbol{{\varphi}}_{A}^{r_u} + \mathbf{t}(\boldsymbol{{\varphi}}_{D}^{r_u})) + \varphi_g}.
    \end{equation}
\end{itemize}
 The effects of unauthorized RIS using directional codebooks will be discussed further in Section \ref{sec:simulation}.
\subsection{Problem Statement for Positioning with Unauthorized RIS}\label{subsec:problem}
Based on the models described above, the positioning problem under interference from an unauthorized RIS is formulated as follows. Given the observations $\mathbf{Y}$ in (1), the first step is to estimate the channel parameter vector $\boldsymbol{\eta} = \begin{bmatrix} \boldsymbol{\eta}_u^\top, \boldsymbol{\eta}_{r_l}^\top  \end{bmatrix}^\top=[\tau_{u}, \rho_{u}, \beta_{u},\phi_{A}^{r_l},\theta_{A}^{r_l},\tau_{r_l}, \rho_{r_l}, \beta_{r_l}]^\top$.
Subsequently, the state vector $\mathbf{s}=[\mathbf{p}_{\rm{U}}^\top, B]^\top$ is determined.\footnote{The complex channel gains for each path are also estimated but treated as nuisance parameters and are not included in the state vector.}  To address RIS-aided positioning in the presence of interference, we derive lower bounds for the channel parameter vector $\boldsymbol{\eta}$ and the state vector $\mathbf{s}$ in Section~\ref{sec:bounds}. In Section~\ref{sec:algorithm}, localization algorithms are proposed to solve this problem.

\section{Bound Analysis}\label{sec:bounds}
We derive the theoretical lower bounds for channel parameters and state parameters to analyze the degradation in positioning performance resulting from the mismatch between the assumed model (without considering unauthorized RIS) and the true model with unauthorized RIS.

\subsection{CRB Analysis for the Interference-free Model}
The CRB \cite{keykhosravi2021siso} for the interference-free model is first derived, which serves as a reference to evaluate the impact of the presence of an unauthorized RIS. 
The channel parameter vectors for the interference-free model and the true model are the same, since the presence of unauthorized RIS will not affect the LOS path and the legitimate RIS path. Based on the signal model in (\ref{eq:mismatched_model}) (\ref{eq:channel_los}) and (\ref{eq:channel_RL}), the fisher information matrix (FIM) $\boldsymbol{\mathcal{I}}( \boldsymbol{\eta}) \in \mathbb{R}^{8 \times 8}$ of the channel estimation is

\begin{equation}
    \boldsymbol{\mathcal{I}}( {\boldsymbol{\eta}}) = \frac{2}{\sigma_n^2} \sum_{k=1}^{K} \sum_{g=1}^{G} \Re \left\{ 
\left( \frac{\partial \tilde{\mu}_{k,g}}{\partial {\boldsymbol{\eta}}} \right)^{\mathrm{H}}  
\left( \frac{\partial \tilde{\mu}_{k,g}}{\partial {\boldsymbol{\eta}}} \right) 
\right\},
\end{equation}
where $\tilde{\mu}_{k,g}$ is the noise-free observation of the received signal $\tilde{\mathbf{Y}} $ in (\ref{eq:mismatched_model}). We then define the CRB on channel parameters as
\begin{equation}
    \text{CRB}_{{\boldsymbol{\eta}}} = \sqrt{\rm{tr}[\boldsymbol{\mathcal{I}}( {\boldsymbol{\eta}})^{-1}]_{1:5,1:5}}.
\end{equation}
The CRB on UE position and clock offset is obtained by
\begin{equation}
    \text{CRB}_{\mathbf{s}} \triangleq \boldsymbol{\mathcal{I}}( \mathbf{s})^{-1} = \left[\mathbf{J}_s \boldsymbol{\mathcal{I}}( {\boldsymbol{\eta}}) \mathbf{J}_s^{\top}\right]^{-1},
\end{equation}
where $\boldsymbol{\mathcal{I}}( \mathbf{s}) \in \mathbb{R}^{4 \times 4}$ is the FIM on the state vector and $\mathbf{J}_s \triangleq \frac{\partial {\boldsymbol{\eta}}}{\partial\mathbf{s} }$ is the Jacobian matrix in a denominator-layout notation.
\subsection{MCRB and MLB Analysis for Channel Parameters}
With the \ac{mcrb} and \ac{mlb} analyzed below, we quantify the performance degradation caused by the presence of an unauthorized RIS in the localization system, which is unaware of its existence. 
We denote $\mathbf{y} \in \mathbb{R}^{GK} = \rm{vec(\mathbf{Y})}$ 
as the received signal vector for the true model (that is, in the presence of the unauthorized RIS). The probability density function (PDF) of the true model can be expressed as $p(\mathbf{y}|\bar{\boldsymbol{\eta}})$, where $\bar{\boldsymbol{\eta}}$ is the vector of true channel parameters that contains all the channel parameters. Similarly, the PDF of the misspecified model is expressed as $\tilde{p}(\mathbf{y}|{\boldsymbol{\eta}})$. 
\subsubsection{Pseudo-True Parameter}
The pseudo-true parameter is the point that minimizes the Kullback-Leibler (KL) divergence between $p(\mathbf{y}|\bar{\boldsymbol{\eta}})$ and $\tilde{p}(\mathbf{y}|{\boldsymbol{\eta}})$. It is defined as \cite{fortunati2017performance}
\begin{equation}
    \boldsymbol{\eta}_0 = \arg\min_{\eta} D_{\rm{KL}}\left(p(\mathbf{y \mid \bar{\boldsymbol{\eta}}}) \,\|\, \tilde{p}(\mathbf{y} \mid \boldsymbol{\eta})\right).
\end{equation}
$\boldsymbol{\mu}(\boldsymbol{\eta})$ and $\tilde{\boldsymbol{\mu}}(\boldsymbol{\eta})$ are the noise-free received signal vectors for the true model and the interference-free model, respectively. The difference between these two vectors is defined as $\boldsymbol{\epsilon}(\boldsymbol{\eta}) \triangleq \boldsymbol{\mu}(\boldsymbol{\eta})-\tilde{\boldsymbol{\mu}}(\boldsymbol{\eta})$. Another interpretation of the pseudo-true parameter $ \boldsymbol{\eta}_0$ is 
 \begin{equation}
\boldsymbol{\eta}_0 = \arg\min_{\boldsymbol{\eta}} \|\boldsymbol{\epsilon}(\boldsymbol{\eta})\|^2.
\end{equation}

\subsubsection{MCRB and MLB }
The theoretical lower bound on the channel parameters in the presence of unauthorized RIS is given as \cite{richmond2015parameter}
\begin{equation}\label{eq:LB_ch}
    \mathrm{MLB}_{\text{ch}}(\boldsymbol{\eta}_0) = \mathrm{MCRB}(\boldsymbol{\eta}_0) + (\bar{\boldsymbol{\eta}} - \boldsymbol{\eta}_0)(\bar{\boldsymbol{\eta}} - \boldsymbol{\eta}_0)^\top.
\end{equation}
Elements in $(\bar{\boldsymbol{\eta}} - \boldsymbol{\eta}_0)(\bar{\boldsymbol{\eta}} - \boldsymbol{\eta}_0)^\top$ in (\ref{eq:LB_ch}) are constant values once the pseudo-true parameters are determined and they are bias terms for lower bounds of channel parameters. 
The covariance matrix of a misspecified-unbiased estimator $\hat{\boldsymbol{\eta}}_{\rm{MS}}$ of $\bar{\boldsymbol{\eta}}$ satisfies
\begin{equation}
   \mathbb{E}_p \left\{ (\hat{\boldsymbol{\eta}}_{\rm{MS}} - \bar{\boldsymbol{\eta}})(\hat{\boldsymbol{\eta}}_{\rm{MS}} - \bar{\boldsymbol{\eta}})^\top \right\} \succeq \mathrm{MLB}(\boldsymbol{\eta}_0), 
\end{equation}
where $\mathbb{E}_p\{\cdot\}$ is the expectation operator. The $\mathrm{MCRB}(\boldsymbol{\eta}_0) \in \mathbb{R}^{8 \times 8}$ can be further elaborated as
\begin{equation}\label{eq:MCRB}
    \mathrm{MCRB}(\boldsymbol{\eta}_0) = \mathbf{A}_{\boldsymbol{\eta}_0}^{-1}\mathbf{B}_{\boldsymbol{\eta}_0} 
 \mathbf{A}_{\boldsymbol{\eta}_0}^{-1},
\end{equation}
where the elements in matrices $\mathbf{A}_{\boldsymbol{\eta}_0}$ and $\mathbf{B}_{\boldsymbol{\eta}_0}$ defined as~\cite{ozturk2023ris}
\begin{equation}\label{eq:An0}
\begin{aligned}
\left[\mathbf{A}_{\boldsymbol{\eta}_0}\right]_{i,j} 
&= \mathbb{E}_p \bigg\{ 
\frac{\partial^2}{\partial \eta_i \partial \eta_j} 
\ln \tilde{p}(\mathbf{y} \mid \boldsymbol{\eta}) 
\bigg|_{\boldsymbol{\eta} = \boldsymbol{\eta}_0}
\bigg\} \\
&= \frac{2}{\sigma_n^2} \Re \bigg[\boldsymbol{\epsilon}(\boldsymbol{\eta})^\herm
\frac{\partial^2 \boldsymbol{\mu}(\boldsymbol{\eta})}{\partial \eta_i \partial \eta_j} 
 \\
&\quad - \left( \frac{\partial \boldsymbol{\mu}(\boldsymbol{\eta})}{\partial \eta_i} \right)^\herm\frac{\partial \boldsymbol{\mu}(\boldsymbol{\eta})}{\partial \eta_j} 
\bigg] \bigg|_{\boldsymbol{\eta} = \boldsymbol{\eta}_0}.
\end{aligned}
\end{equation}
\begin{equation}
\begin{aligned}
\left[\mathbf{B}_{\boldsymbol{\eta}_0}\right]_{i,j} &= 
\mathbb{E}_p \left\{
\frac{\partial}{\partial \eta_i} \ln \tilde{p}(\mathbf{y} \mid \boldsymbol{\eta}) 
\frac{\partial}{\partial \eta_j} \ln \tilde{p}(\mathbf{y} \mid \boldsymbol{\eta}) 
\bigg|_{\boldsymbol{\eta} = \boldsymbol{\eta}_0}
\right\} \\
&= \frac{4}{\sigma_n^4} \Re \left[
\boldsymbol{\epsilon}(\boldsymbol{\eta})^\herm
\frac{\partial \boldsymbol{\mu}(\boldsymbol{\eta})}{\partial \eta_i} 
\right]
\Re \left[
\boldsymbol{\epsilon}(\boldsymbol{\eta})^\herm
\frac{\partial \boldsymbol{\mu}(\boldsymbol{\eta})}{\partial \eta_j}
\right] \\
&\quad + \frac{2}{\sigma_n^2} \Re \left[
\left( \frac{\partial \boldsymbol{\mu}(\boldsymbol{\eta})}{\partial \eta_i} \right)^{\herm}
\frac{\partial \boldsymbol{\mu}(\boldsymbol{\eta})}{\partial \eta_j}
\right] \bigg|_{\boldsymbol{\eta} = \boldsymbol{\eta}_0}.
\end{aligned}
\label{eq:Bn0}
\end{equation}

\subsection{Absolute Lower Bound for 3D positioning}\label{subsec:ALB}
Estimation of position and clock offset based on estimated channel parameter vectors in the presence of an unauthorized RIS in the system is also a mismatched estimation problem. The lower bound for this estimation is given by
\begin{equation}
       \mathrm{MLB}_{\rm{pos}}(\mathbf{s}_0) = \mathrm{MCRB}(\mathbf{s}_0) + (\bar{\mathbf{s}} - \mathbf{s}_0)(\bar{\mathbf{s}} - \mathbf{s}_0)^\top, 
\end{equation}
where $\bar{\mathbf{s}}$ and $\mathbf{s}_0)$ are the true and the pseudo-true state parameter vectors, respectively. $\mathbf{s}_0$ can be obtained by~\cite{chen2023modeling}
\begin{equation}
\mathbf{s}_0 = \arg\min_{\bar{s}} 
\left( \boldsymbol{\eta}_{0} - \boldsymbol{\eta}(\bar{\mathbf{s}}) \right)^\top 
\boldsymbol{\mathcal{I}}(\boldsymbol{\eta}) 
\left( \boldsymbol{\eta}_{0} - \boldsymbol{\eta}(\bar{\mathbf{s}}) \right).
\end{equation}
The MCRB for the state parameter vector can, in principle, be derived using calculations similar to those in (\ref{eq:MCRB}), (\ref{eq:An0}), and (\ref{eq:Bn0}). However, the computational complexity of this derivation becomes significantly high, especially when a 3D scenario and massive transmissions are considered. Additionally, the lower bound for the state parameter vector converges to the bias term, given by
$(\bar{\mathbf{s}} - \mathbf{s}_0)(\bar{\mathbf{s}} - \mathbf{s}_0)^\top.$
As a result, our analysis focuses primarily on the bias component of the lower bound of the state parameter in the subsequent discussion. This component is further defined as $\text{ALB}_{\rm{pos}} = (\bar{\mathbf{s}} - \mathbf{s}_0)(\bar{\mathbf{s}} - \mathbf{s}_0)^\top.$

\section{Localization Algorithms}\label{sec:algorithm}
In this section, A robust localization algorithm is proposed, incorporating a low-complexity method to obtain coarse estimation results, followed by MLE to achieve refined estimation results. These algorithms aim to address the problem formulated in Section \ref{subsec:problem} and consist of two main steps. The first step focuses on estimating the channel parameters, while the second step determines the position of the UE and the clock offset of the BS. For convenience of algorithm development later, here we define the nuisance-free channel parameter vector as $\boldsymbol{\eta}_{N} = [\tau_{u}, \phi_{A}^{r_l},\theta_{A}^{r_l},\tau_{r_l}]^\top$.
\subsection{Low-Complexity Estimation}
In the absence of interference from an unauthorized RIS, the received signals from the LOS and legitimate RIS paths can be effectively separated employing an orthogonal RIS profile on the legitimate RIS, as shown in (\ref{eq:los_mismatched}) and (\ref{eq:ris_mismatched}). In this case, the delays of the LOS and legitimate RIS paths can be estimated by applying a discrete Fourier transform (DFT) to the corresponding channel frequency responses, as demonstrated in \cite{chen2024multi}. However, in the actual signal model, interference from unauthorized RIS degrades the path separation performance, as shown in (\ref{eq:updated_yu}) and (\ref{eq:updated_yrl}). The following section presents an approach to estimating delays in the presence of an unauthorized RIS.  


\subsubsection{LOS Channel Parameter Estimation}  
For the LOS path, interference from unauthorized RIS has minimal impact on the received signals, as the LOS path signal power is significantly higher than that of the unauthorized RIS path. The latter experiences a greater propagation loss due to its longer link distance, an effect that is particularly pronounced in the mmWave and higher frequency bands \cite{rappaport2015wideband}. Consequently, the LOS path delay can still be accurately estimated by applying a DFT to the estimated channel response of the LOS path.  
To this end, we first estimate the channel response vector of LOS path $\hat{\mathbf{h}}_u\in \mathbb{R}^K$ as
\begin{equation}
\hat{\mathbf{h}}_u = \sum_{g=1}^{{G/2}} \left[\mathring{\mathbf{Y}}_u\right]_{:,g} \odot \mathbf{x}^*.
\end{equation}
The estimated delay of the LOS path $\hat{\tau}_u$ is then obtained by
\begin{equation}\label{eq:tau_u_est}
\hat{\tau}_u = \arg\max_{\tau} \left| \mathbf{d}^{\mathrm{H}}(\tau) \hat{\mathbf{h}}_u \right|.
\end{equation}

\subsubsection{RIS Channel Parameter Estimation}
Estimating the delay of the legitimate RIS path is more challenging due to interference from the unauthorized RIS. The estimated channel response for the legitimate RIS path is given by
\begin{equation}
\hat{\mathbf{H}}_{\rm{R_L}} = \mathring{\mathbf{Y}}_{\rm{R_L}} \odot \left(\mathbf{x}^* \mathbf{1}_{G/2}^\top \right) \in \mathbb{R}^{K \times G/2}.
\end{equation}
If we apply a method similar to the one shown in \eqref{eq:tau_u_est}, the estimated delay for the legitimate RIS is given by
\begin{equation}\label{eq:tau_est}
    \hat{\tau}_{r_l} = \arg\max_{\tau} \| \mathbf{d}^{\mathsf{H}}(\tau) \widehat{\mathbf{H}}_{\rm{R_L}} \|.
\end{equation}
\begin{figure}
    \centering    \includegraphics[width=0.95\linewidth]{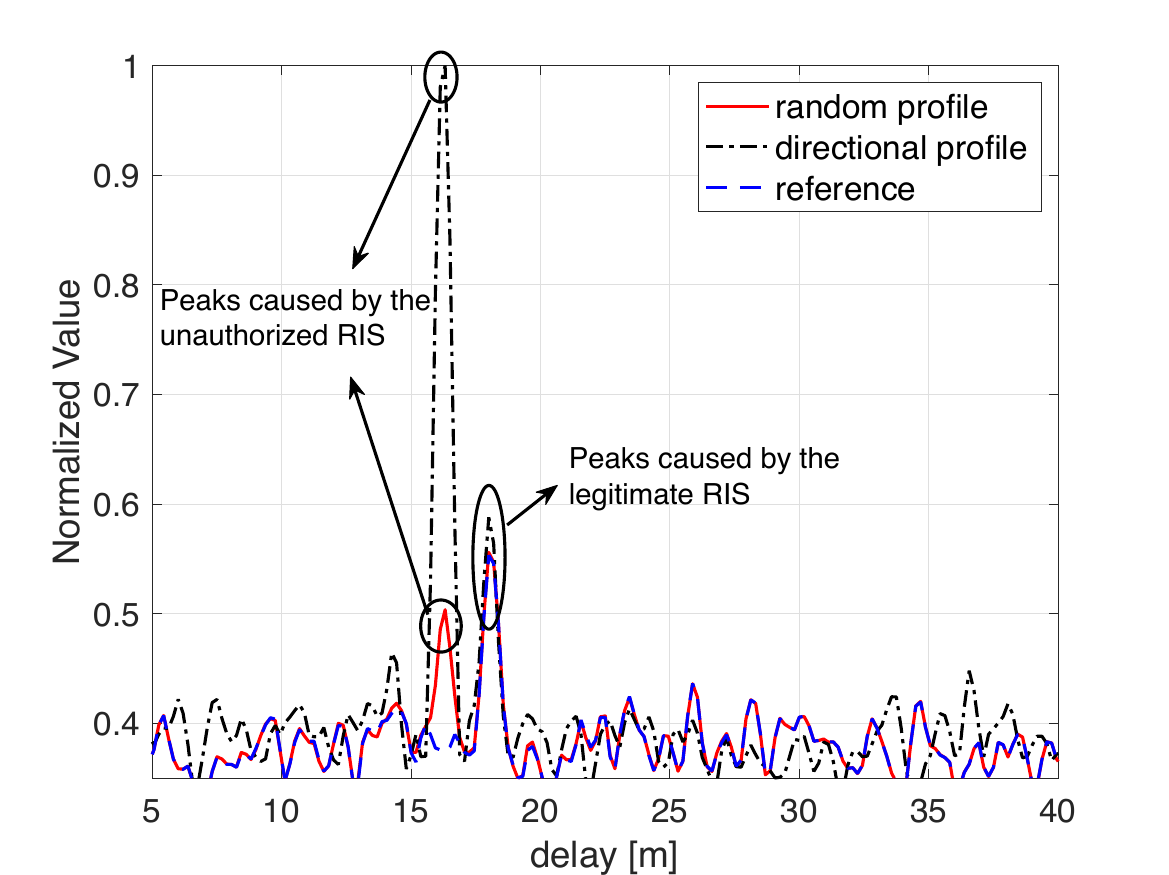}
    \caption{DFT results for estimating delay of legitimate RIS path with different beamforming codebooks (i.e., random and RPDC) for unauthorized RIS. (Note that other simulation settings are the same as the settings in Section~\ref{sec:simulation}).}
\label{fig:fft_results}
\end{figure}
However, directly applying a DFT to \( \hat{\mathbf{H}}_{\rm{R_L}} \) and selecting the delay corresponding to the maximum value as implemented in~\cite{chen2024multi} is unlikely to yield the correct delay estimate. As shown in Fig.~\ref{fig:fft_results}, the presence of the unauthorized RIS introduces multiple significant peaks, whereas only a single dominant peak appears in the reference case without interference. When comparing results obtained using a random codebook and a directional codebook (RPDC) for the unauthorized RIS, the interference-induced peak can even surpass the legitimate RIS peak, leading to incorrect delay estimation if the LOS path method is applied. To address this issue, we identify all prominent peaks in the DFT result for \( \hat{\mathbf{H}}_{\rm{R_L}} \) using a one-dimensional search.  

Assuming a total of $I$ local maxima are identified, the estimated delay vector consists of the delay bins corresponding to these recognized peaks, expressed as $\boldsymbol{\widehat{\mathbf{\tau}}}=[\widehat{\tau}_1,\cdots,\widehat{\tau}_I]^\top$. The estimated channel parameters of the legitimate RIS, \( \hat{\phi}_{A}^{r_l}, \hat{\theta}_{A}^{r_l}, \) and \( \hat{\tau}_{r_l} \), is then obtained by solving 
\begin{equation}\label{eq:ch_solve}
    \left[\hat{\phi}_{A}^{r_l}, \hat{\theta}_{A}^{r_l}, \hat{\tau}_{r_l}\right]  = \arg\max_{\phi, \theta,\tau \in \boldsymbol{\widehat{\mathbf{\tau}}}} \mathcal{L}_{g,k} (\phi, \theta,\tau),
\end{equation}
where 
\begin{equation*}
    \mathcal{L}_{g,k} (\phi, \theta,\tau) = \sum_{g,k} \bigg| \omega_g^{r_l} e^{\frac{j2\pi}{\lambda_c} \mathbf{Z}^\top \mathbf{t}(\boldsymbol{\varphi})}d_k({\tau}) x_k \mathring{y}^*_{k,g} \bigg|.
\end{equation*}
Moreover, $\mathbf{t}(\boldsymbol{\varphi})$ is derived from $\phi$ and $\theta$ according to (\ref{eq:dir_vec_rl}) and $\mathring{y}_{k,g}$ is the entry of the matrix $\mathring{\mathbf{Y}}$.
A 2D search with a specified angular step size is performed for $\phi$ and $\theta$ to solve the problem described in (\ref{eq:ch_solve}).

\subsubsection{Closed-form Position Estimation}
We next derive a closed-form solution for state vector estimation, based on the estimated \ac{aoa} pairs $\hat{\phi}_{A}^{r_l}$, $\hat{\theta}_{A}^{r_l}$ and delays $\hat{\tau}_{u}$, $\hat{\tau}_{r_l}$. 
The key idea is to determine the UE position along the direction vector from the legitimate RIS, given by $\hat\tv = \tv(\hat{\phi}_{A}^{r_l}, \hat{\theta}_{A}^{r_l})$, as shown in~\eqref{eq:dir_vec_rl}. To facilitate problem formulation, we define $d_R \triangleq \Vert \pv_{\rm{B}} - \pv_{\rm{R_L}}\Vert$ as the distance between the legitimate RIS and BS. Additionally, we introduce an auxiliary parameter $r$ as the distance between the UE and legitimate RIS, such that 
\begin{equation}
    \hat\pv_{\rm{U}} (r) = \pv_{\rm{R_L}} + r\hat \tv.
    \label{eq_hat_pu}
\end{equation}
Subtracting~\eqref{eq:tau_ris_l} from~\eqref{eq:tau_los}, the clock offset is canceled out as 
\begin{equation}
    r + \underbrace{d_{\rm{R}} + c\hat\tau_{u}- c\hat\tau_{r_l}}_{\hat d_\Delta} 
    = \Vert \pv_{\rm{U}} - \pv_{\rm{B}} \Vert.
\end{equation}
After getting the squared of both sides we obtain
\begin{equation}
    (r + \hat d_\Delta)^2 = r^2 + 2r\hat \tv^\top(\pv_{\rm{U}} - \pv_{\rm{B}}) + d^2,
\end{equation}
and the closed-form solution of the auxiliary parameter $r$ is
\begin{equation}
    r = \frac{d^2 - \hat d_\Delta^2}{2\hat \tv^\top(\pv_{\rm{U}} - \pv_{\rm{B}}) - 2\hat d_\Delta}.
\end{equation}
The estimated parameter $\hat r$ is then substituted in~\eqref{eq_hat_pu} to obtain $\hat\pv_{\rm{U}}$, while $B$ can be estimated accordingly. In interference-free scenarios, a \ac{mle} can be employed to achieve higher positioning accuracy~\cite{keykhosravi2022ris}. The output of the low-complexity estimator can serve as an initial point for \ac{mle}, followed by iterative gradient-based refinement.  

\subsection{MLE}
MLE can be used to estimate the optimal channel parameters. However, the true received signal vector $\mathbf{y}$ is observed in the presence of the unauthorized RIS in our case, and the MLE does not know the signal model of the unauthorized RIS path. Therefore, there is a mismatch between the signal model assumed by the estimator and the true signal model. The channel parameters estimated by the MLE can be given as
\begin{equation}\label{eq:MMLE_concept}
\hat{\boldsymbol{\eta}}_{\text{MLE}} = \arg\max_{\boldsymbol{\eta}} \ln \tilde{p}(\mathbf{y}|{\boldsymbol{\eta}}),
\end{equation}
where $\ln \tilde{p}(\mathbf{y}|{\boldsymbol{\eta}})$ is the log-likelihood of the mismatched model~\eqref{eq:mismatched_model}.
This MLE estimator can still be used to refine the estimation results. The estimated nuisance-free channel parameters obtained from the first step of the low-complexity estimator can be used as an initial estimate for the following refining estimation process. Based on (\ref{eq:MMLE_concept}), the refined estimation is  formulated as
\begin{equation}
\hat{\boldsymbol{\eta}} = \arg\min_{\boldsymbol{\eta}} \left\| \mathbf{y} - \boldsymbol{\mu}(\boldsymbol{\eta}) \right\|^2.
\end{equation}
To further refine the initial estimate of the nuisance-free channel parameters, the estimation is optimized as follows \cite{chen2024multi}.
\begin{equation}\label{eq:MMLE_final}
\hat{\boldsymbol{\eta}}_{N} = \arg\min_{\boldsymbol{\eta}_{N}} 
\left\| 
\mathring{\boldsymbol{\mu}} - 
\frac{\boldsymbol{\mu}^\mathrm{H}(\boldsymbol{\eta}_{N}) \mathring{\boldsymbol{\mu}}}
{\|\boldsymbol{\mu}(\boldsymbol{\eta}_{N})\|^2} 
\boldsymbol{\mu}(\boldsymbol{\eta}_{N})
\right\|,
\end{equation}
where $\mathring{\boldsymbol{\mu}} = \text{vec}(\mathring{\mathbf{Y}}_u+\mathring{\mathbf{Y}}_{\rm{R_L}})$ and $\boldsymbol{\mu}(\boldsymbol{\eta}_{N}) = (\mathbf{d}(\tau_u)\mathbf{1}_{G}^{\mathrm{T}}+(\mathbf{d}(\tau_{r_l})\mathbf{1}_{G}^{\mathrm{T}}) \odot \mathbf{A}(\boldsymbol{\varphi}_{D}^{r_u}, \boldsymbol{\varphi}_{A}^{r_u}))\odot \mathbf{X}$. Note that the nuisance channel parameters \( \alpha_u \) and \( \alpha_{r_l} \), which are constant values, are removed in (\ref{eq:MMLE_final}).  

With the refined nuisance-free channel parameters and the initial state parameters obtained in~\eqref{eq:MMLE_final}, the refined state parameter vector using mismatched MLE is obtained as
\begin{equation}
\hat{\mathbf{s}} = \arg\min_{\mathbf{s}} 
\left( \hat{\boldsymbol{\eta}}_N - \boldsymbol{\eta}_N(\mathbf{s}) \right)^\top 
 \boldsymbol{\mathcal{I}}( {\boldsymbol{\eta}}_N) 
\left( \hat{\boldsymbol{\eta}}_N - \boldsymbol{\eta}_N(\mathbf{s}) \right),
\end{equation}
where $ \boldsymbol{\mathcal{I}}( {\boldsymbol{\eta}}_N) \in \mathbb{R}^{4 \times 4}$ is the equivalent FIM which can be simply obtained from the submatrix of $ \boldsymbol{\mathcal{I}}( {\boldsymbol{\eta}})$.

\section{Numerical Simulation} \label{sec:simulation}
We consider a 3D RIS-aided positioning scenario involving one legitimate RIS and one unauthorized RIS, as illustrated in Fig.~\ref{fig:system}. The default simulation parameters are summarized in Table~\ref{tab:settings}.  A total of $K = 512$ subcarriers are utilized as pilot signals for positioning. The transmitted symbol vector $\mathbf{x}$ consists of random unit-modulus signals, where each element has a unit amplitude and a phase uniformly distributed in the range $[0, 2\pi)$. Unless stated otherwise, the legitimate RIS employs a random codebook. 
\begin{table}[ht]
\centering
\caption{Default Parameter Settings}
\begin{tabular}{|l|l|}
\hline
\multicolumn{2}{|c|}{\textbf{Signal \& Frequency Parameters}} \\ \hline
Center frequency                  & 30 GHz      \\ \hline
Bandwidth                         & 400 MHz        \\ \hline
Number of subcarriers             & $K$ = 256 \\ \hline
Number of symbols                 & $G$ = 192    \\ \hline
Noise figure                      & 10 dB    \\ \hline
clock offset                      & $B = 5$ m    \\ \hline
\multicolumn{2}{|c|}{\textbf{Geometry Parameters}} \\ \hline
BS position               &  $\pv_{\rm B} = [0, 0, 0]\top$   \\ \hline
UE position               & $\pv_{\rm U} = [1, 4, -2]\top$     \\ \hline
Dimensions of legitimate RIS       & $N = 10 \times 10$         \\ \hline
Dimensions of unauthorized RIS     & $M = 10 \times 10$          \\ \hline
Legitimate RIS position    & $\pv_{\rm {R_L}} = [-2, 4, 0]\top$      \\ \hline
Legitimate RIS orientation & $\ov_{\rm {R_L}} = [0, 0, 0]\top$         \\ \hline
Unauthorized RIS position  & $\pv_{\rm {R_U}} = [2, 5, 0]\top$        \\ \hline
Unauthorized RIS orientation & $\ov_{\rm {R_U}} = [0, 0, 0]\top$         \\ \hline
\end{tabular}
\label{tab:settings}
\end{table}
\subsection{Bounds and Estimation Results}
\begin{figure}
    \centering
    \begin{minipage}[t]{0.46\textwidth} 
        \centering
        \includegraphics[width=\textwidth]{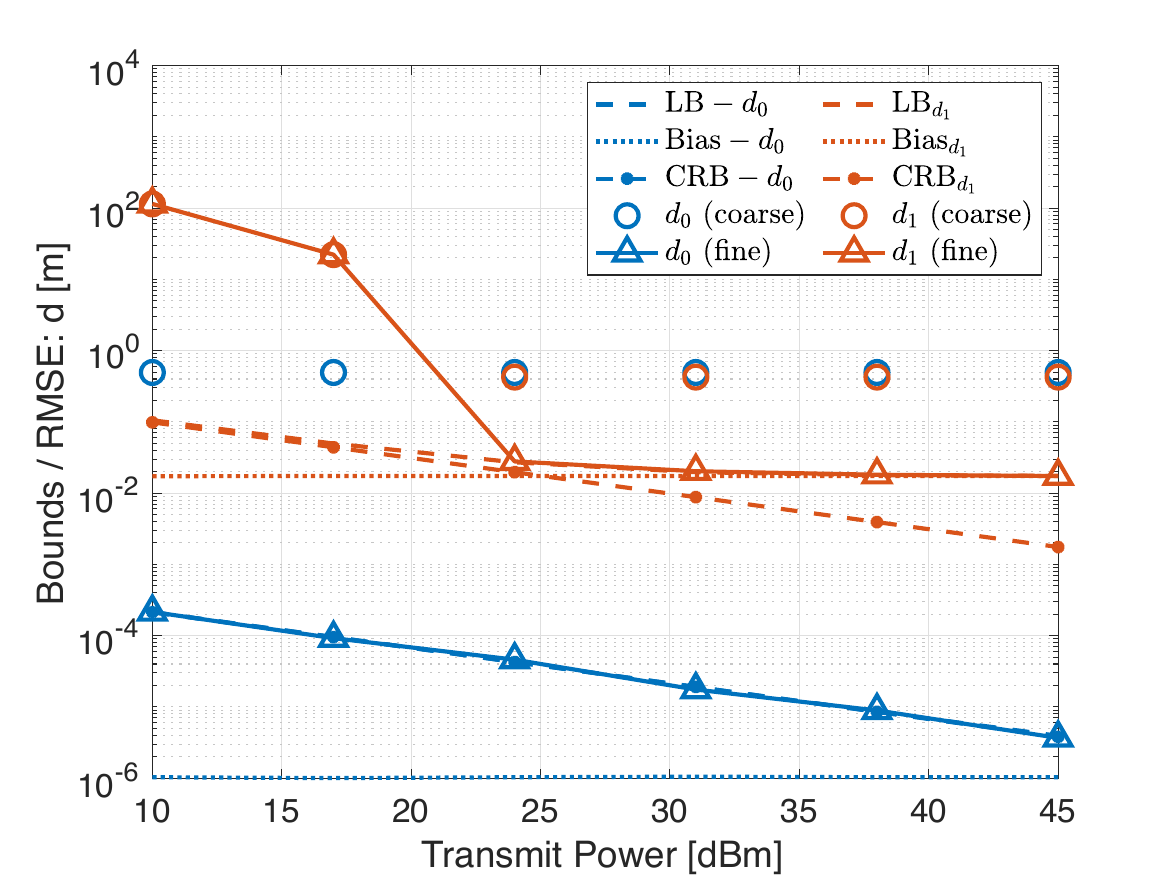} {\small{
        \*{(a)}}}
    \end{minipage}
    \hfill
    \begin{minipage}[t]{0.46\textwidth}
        \centering
        \includegraphics[width=\textwidth]{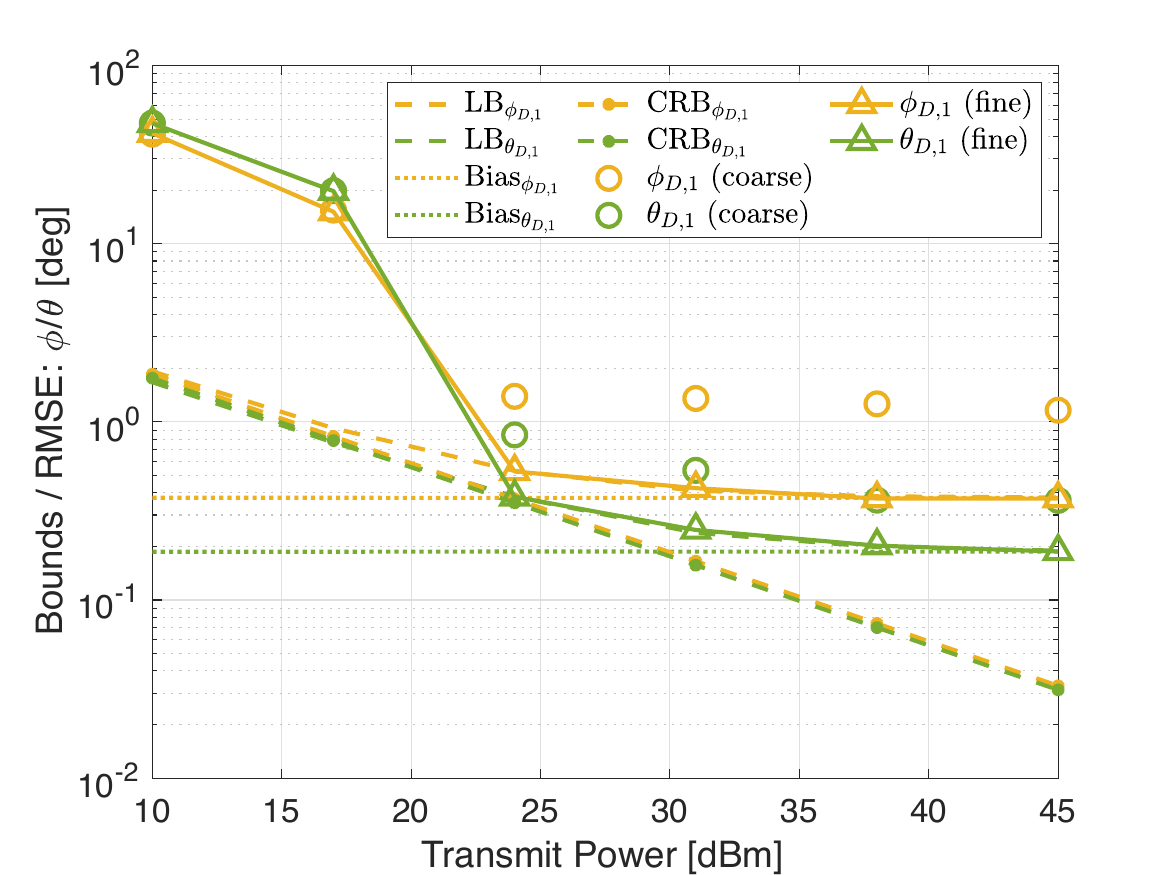}
        {\small{
        \*{(b)}}}
    \end{minipage}
    \caption{Different theoretical bounds and RMSE of the estimation results for channel parameters: (a) delays of LOS and legitimated RIS paths; (b) AOA of the legitimate RIS path. Both the legitimate and unauthorized RIS employ the random codebook.}
   \label{fig:ch_results}
\end{figure}
\begin{figure}
    \centering
    \includegraphics[width=0.95\linewidth]{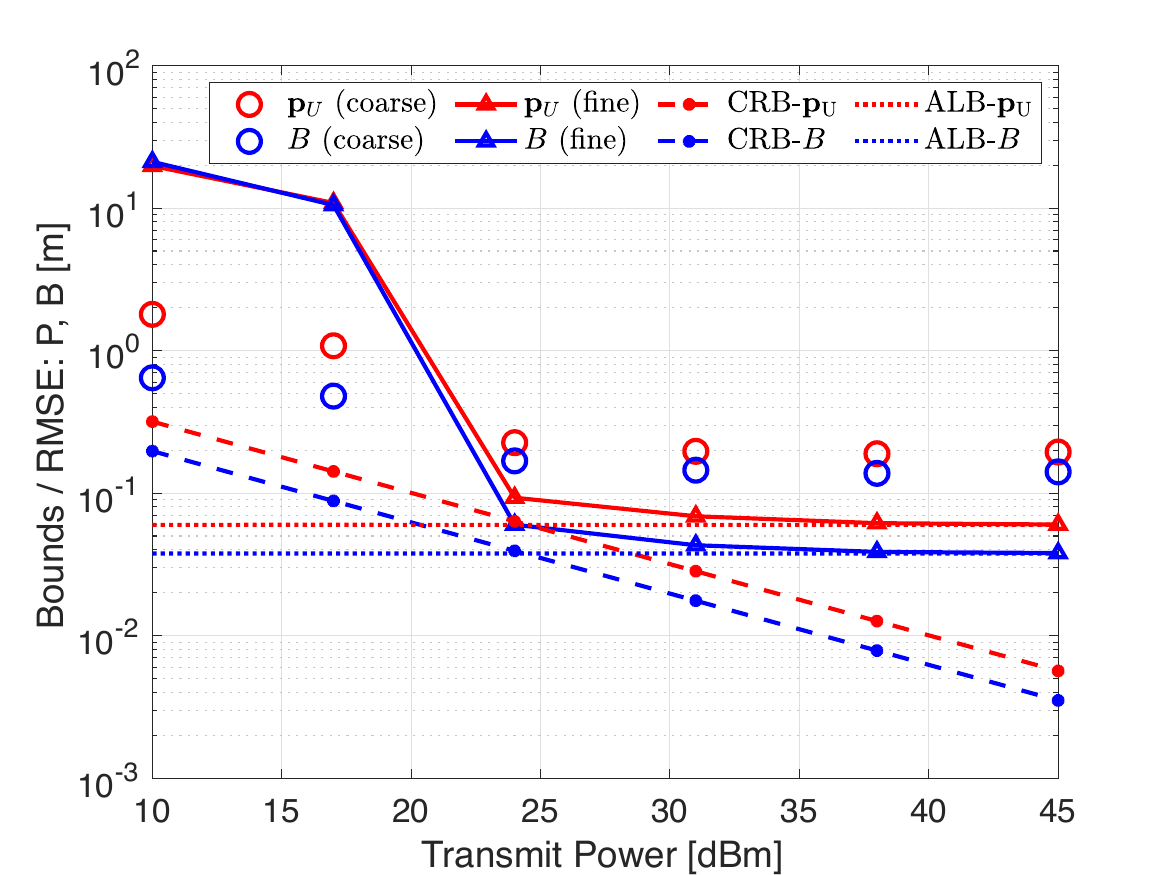}
    \caption{Different theoretical bounds and RMSE of the estimation results for state parameters Both the legitimate and unauthorized RIS employ the random codebook. }
    \label{fig:loc_results}
\end{figure}
For the low-complexity estimation, the resolution of the grid search for $\phi$ and $\theta$ in (\ref{eq:ch_solve}) is set to $1.6^\circ$. The search area around the true UE position, as mentioned in (\ref{eq:ch_solve}), is defined as a $2 \times 2 \times 2\ \text{m}^3$ space with a step size of $0.2$ m in each of the three dimensions. The derived bounds and RMSE of the estimated channel parameters, with both the legitimate and unauthorized RIS employing a random codebook, are presented in Fig.~\ref{fig:ch_results}. The coarse estimation represent the results obtained from the low-complexity estimation process, while the fine results refer to the estimations refined through MLE. The CRBs of the interference-free model, the MLBs as well as the bias of the corresponding MLBs serve as the benchmarks for estimators. Note that the impact of different beamforming codebooks for the unauthorized RIS will be analyzed in subsequent discussions.  As shown in Fig.~ \ref{fig:ch_results} (a), the delay result for the LOS path is particularly different from the other parameters. The CRB and MLB for this parameter are nearly identical, with the bias of its MLB being less than \(10^{-6}\) m, which can be considered negligible. This indicates that the presence of an unauthorized RIS has a minimal impact on the LOS path, as the LOS path is considerably stronger than the path associated with the unauthorized RIS. The RMSE for the coarse estimation result remains almost unchanged as the transmit power increases, while the fine estimation result consistently aligns with the CRB and MLB. For the delay of the legitimate RIS path, the difference between its MLB and the CRB becomes apparent when the transmit power exceeds 25 dBm, as the MLB gradually converges toward the bias term. The coarse estimation result saturates at a level higher than the MLB, whereas the refined estimation result aligns with the MLB when the transmit power exceeds 24 dBm. 

The bounds and estimation results for the AOA of the legitimate RIS path are shown in Fig.~\ref{fig:ch_results} (b). A similar trend can be observed for the bounds of $\phi$ and $\theta$, while the bias for $\phi$ is a bit larger than that for $\theta$ at this specific location. It can be inferred that the unauthorized RIS has minimal impact on the performance of the estimators when the transmit power is low, as the performance is primarily limited by noise effects at this stage. However, as the transmit power increases, the bias term of the MLB becomes dominant, ultimately setting the lower bound on performance. Moreover, with varying relative positions of the legitimate RIS and the unauthorized RIS, interference from the unauthorized RIS can have different impacts on the legitimate RIS path, leading to relatively higher or lower MLB and bias values in terms of delay and AOA. The results in Fig. \ref{fig:ch_results} demonstrate the effectiveness of the localization algorithms and the derived bounds. The choice between the low-complexity estimator and the MLE depends on the specific requirements for computational burden and estimation accuracy.     

The theoretical bounds and the estimation results for the UE position and system clock offset are shown in Fig.~\ref{fig:loc_results}. The CRB bounds (dashed lines with circle markers) represent the theoretical lower bounds for positioning accuracy in the absence of an unauthorized RIS. The ALBs for the UE position and clock offset correspond to the bias terms of the MLBs when interference from an unauthorized RIS is present, as discussed in Section~\ref{subsec:ALB}.  
At low transmit power levels, the coarse estimation results exhibit smaller estimation errors compared to the refined estimation results. This occurs because the grid search used in coarse estimation is confined to a $2 \times 2 \times 2 \, \mathrm{m}^3$ region, providing a constrained search space that improves robustness under low transmit power conditions. In contrast, the optimization process in refined estimation is not restricted to this area, making its advantages less evident at lower power levels. However, as the transmit power increases beyond 24 dBm, the refined estimation results for the state parameters converge toward the ALBs, which ultimately serve as their lower bound.  

\begin{figure}
    \centering
    \includegraphics[width=0.95\linewidth]{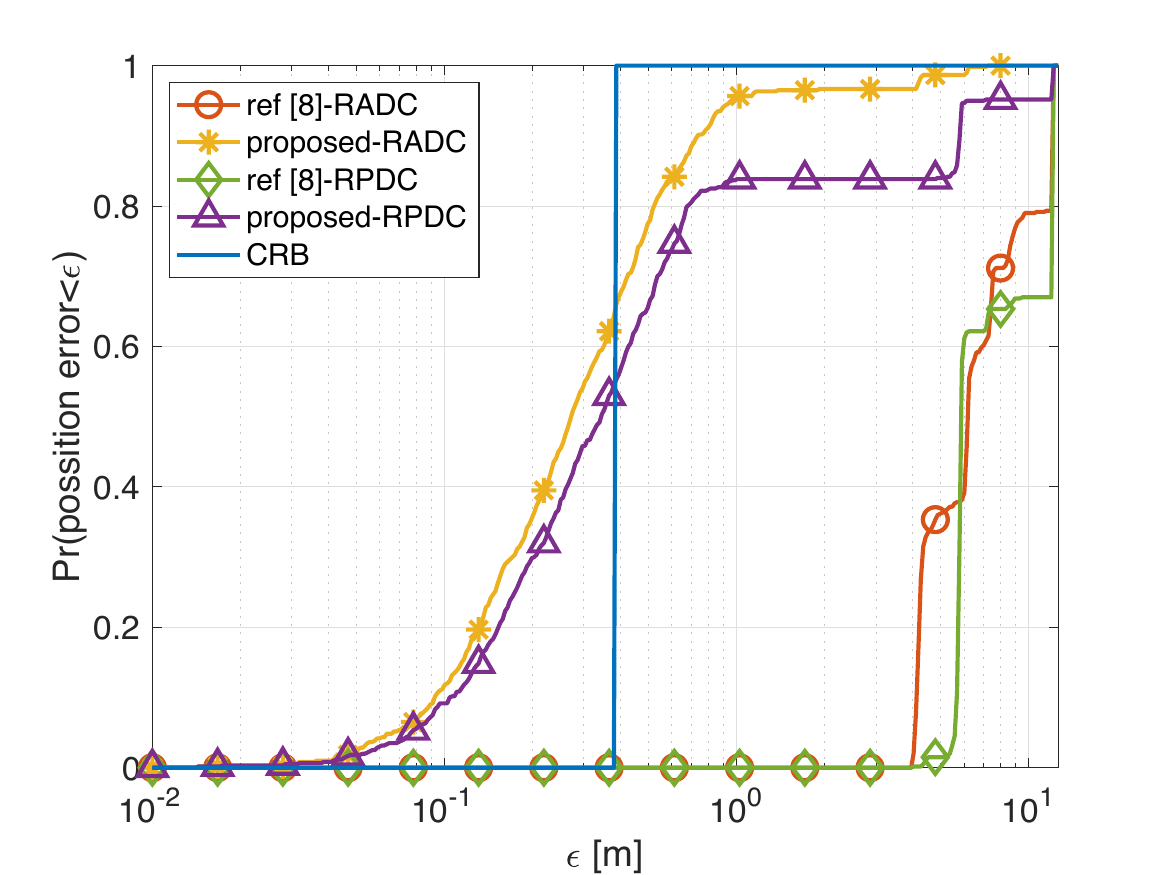}
    \caption{Comparison of the CDF of CRB and RMSE for $\pv_{\rm U}$ using different localization algorithms, with the unauthorized RIS employing RADC and RPDC. }
    \label{fig:estimation_CDF}
\end{figure}
As mentioned in Section \ref{sec:algorithm}-A, the use of RADC and RPDC could introduce challenges on estimating the legitimate RIS path parameters. Using the algorithm proposed in \cite{chen2024multi} could mistake the peak caused by the unauthorized RIS (as shown in Fig. \ref{fig:fft_results}) and obtain large errors on RIS path channel estimation. The CDF results of RMSE in terms of $\pv_{\rm U}$ using the algorithm in \cite{chen2024multi} and the proposed one are compared in \ref{fig:estimation_CDF} to evaluate the estimator performance when the unauthorized RIS employs RPDC. The CDF of CRB here is considered as the result without interference. As observed, the proposed algorithm can achieve much better positioning performance than the method in the literature when RADC or RPDC is used for unauthorized RIS. It can also be observed that the unauthorized RIS using RPDC creates a more adverse condition compared to RADC, as it is more likely to result in higher positioning estimation errors.

\subsection{The Effect of Beamforming Codebook}
\begin{figure}
    \centering
    \begin{minipage}[t]{0.46\textwidth} 
        \centering
        \includegraphics[width=\textwidth]{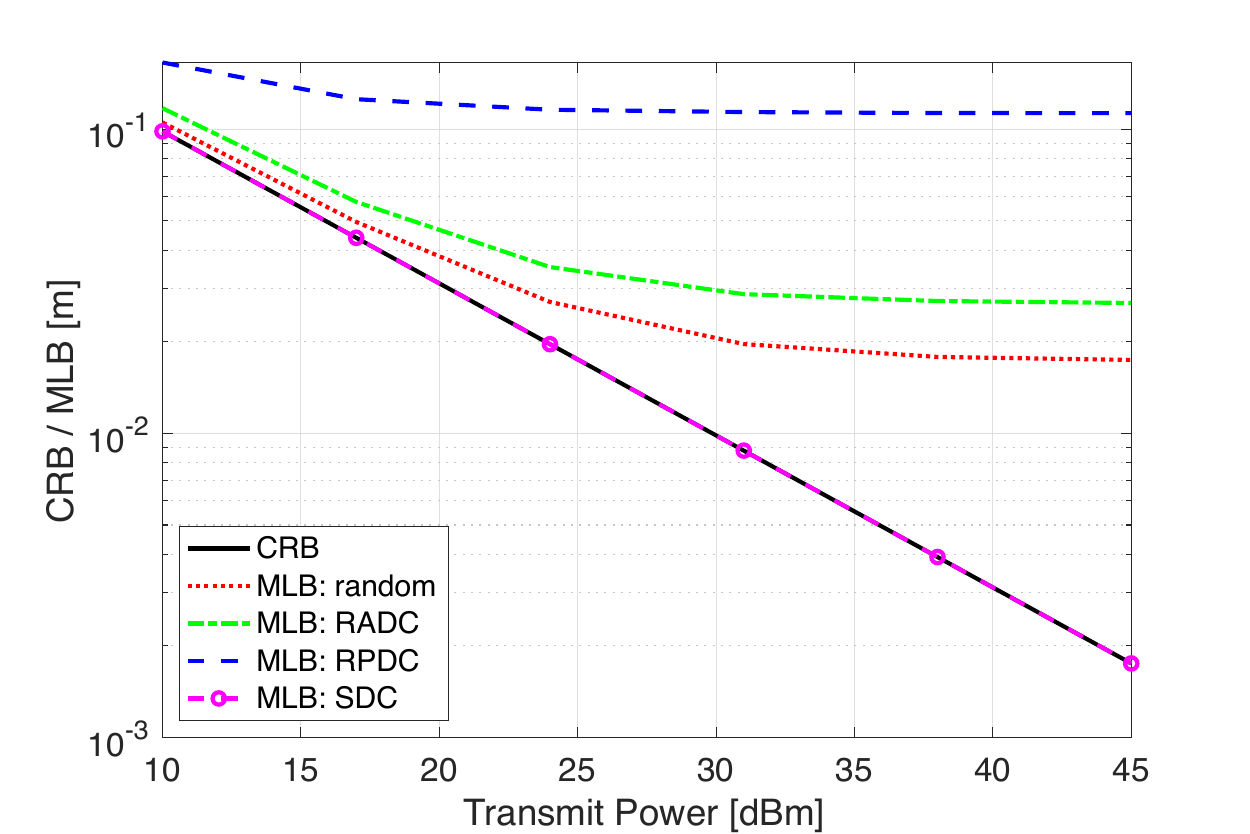} {\small{
        \*{(a)}}}
    \end{minipage}
    \hfill
    \begin{minipage}[t]{0.46\textwidth}
        \centering
        \includegraphics[width=\textwidth]{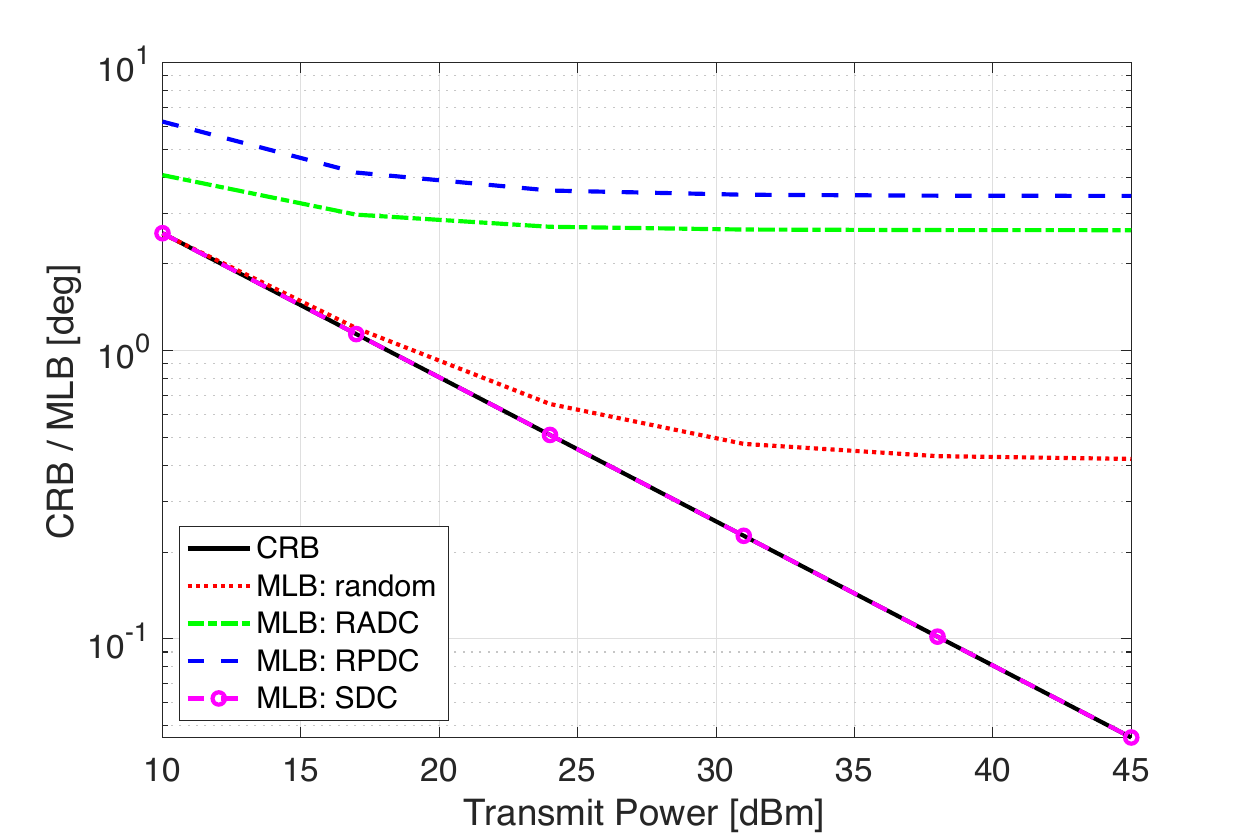}
        {\small{
        \*{(b)}}}
    \end{minipage}
    \caption{CRB and MLB results with different unauthorized RIS beamforming codebooks in terms of (a) the delay (b) AOA: $\boldsymbol{\varphi}$ of the legitimate RIS path, respectively.  }
    \label{fig:ris_profiles}
\end{figure}
We assume that the legitimate RIS employs a random beamforming codebook. In this scenario, different beamforming codebooks are considered for the unauthorized RIS to evaluate their impact on channel estimation performance. These include a random codebook, SDC (derived  from(\ref{eq:codebook_dir0})), RADC (derived from (\ref{eq:codebook_dir1})), and RPDC (derived from (\ref{eq:codebook_dir2})). For RADC, the AOA uncertainties in (\ref{eq:codebook_dir1}), from inaccurate prior knowledge of the UE, are set as $\Delta \phi_g \sim \mathcal{U}[-10^\circ, 10^\circ]$, $\Delta \theta_g \sim \mathcal{U}[-5^\circ, 5^\circ]$.  

The CRB and MLB results for the channel parameters of the legitimate RIS path, using different unauthorized RIS beamforming codebooks, are compared in Fig.~\ref{fig:ris_profiles}\footnote{The bounds for \( \boldsymbol{\varphi} \) are obtained by taking the square root of the sum of the results for \( \phi \) and \( \theta \).}. The CRB results provide a benchmark for performance in the absence of an unauthorized RIS. It is surprisingly observed that the MLB of results for SDC aligns with the CRB results, indicating unsuccessful interference. The SDC, detailed in \eqref{eq:codebook_dir0}, remains unchanged as $g$ varies. Consequently, the legitimate RIS path could effectively be separated from the received signals as $[\mathbf{Y_{\rm{R_U}}}]_{:,k} = [\mathbf{Y_{\rm{R_U}}}]_{:,k+\frac{G}{2}}$. As a result, the localization system experiences negligible interference from the unauthorized RIS. Besides, it can be observed that, at low transmit power, the MLB and CRB exhibit minimal differences when a random beamforming codebook is used, indicating a negligible impact from the unauthorized RIS in this case. However, for the two directional codebooks, higher MLB values are observed across all legitimate RIS channel parameters compared to those obtained with the random codebook.  
This indicates that employing directional codebooks on the unauthorized RIS can generate stronger interference compared to using a random codebook. 
When comparing the results for the two directional codebooks, a higher MLB is observed when using RPDC for legitimate RIS path parameters. This indicates that RPDC induces stronger and more effective interference compared to RADC.
\begin{figure*}[ht!]
    \centering
    \begin{minipage}[t]{0.32\textwidth}
        \centering
        \includegraphics[width=\textwidth]{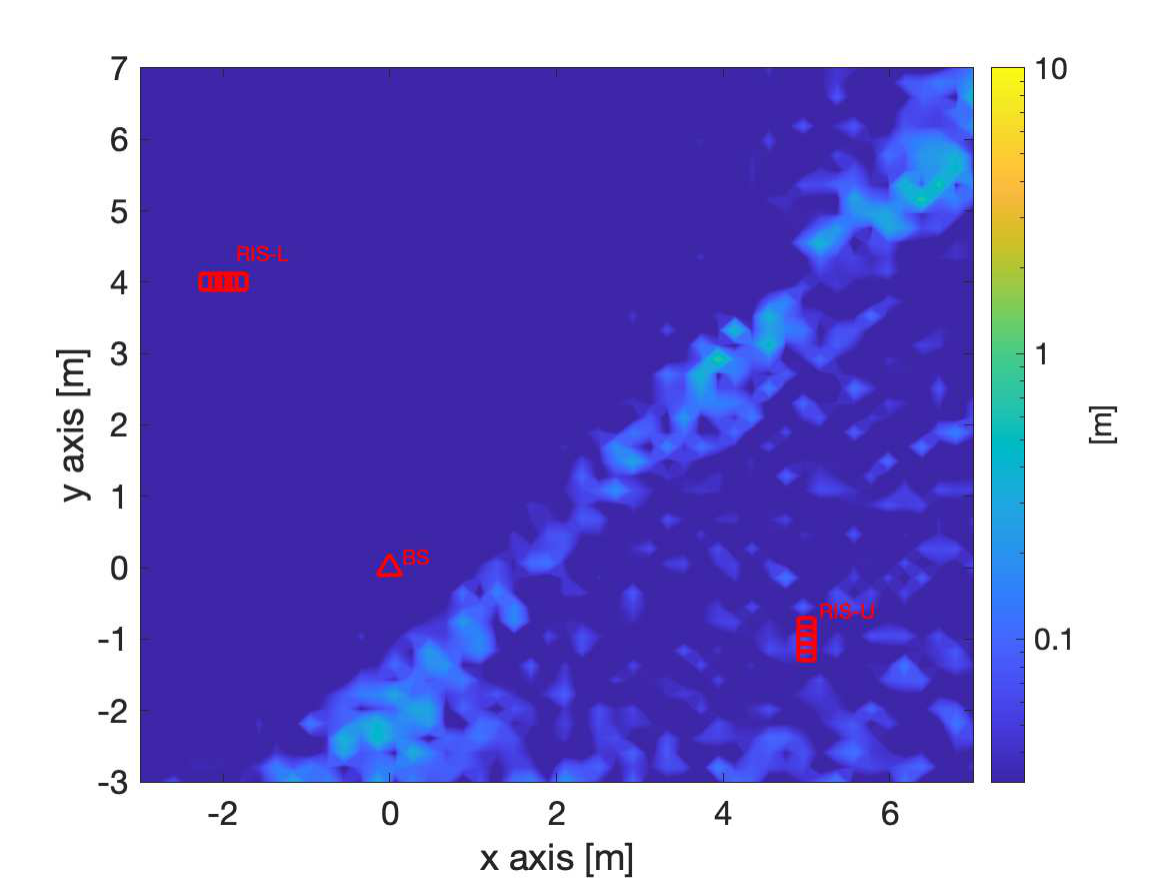} 
        \par Scenario 1 - random
    \end{minipage}
    \hfill
    \begin{minipage}[t]{0.32\textwidth}
        \centering
        \includegraphics[width=\textwidth]{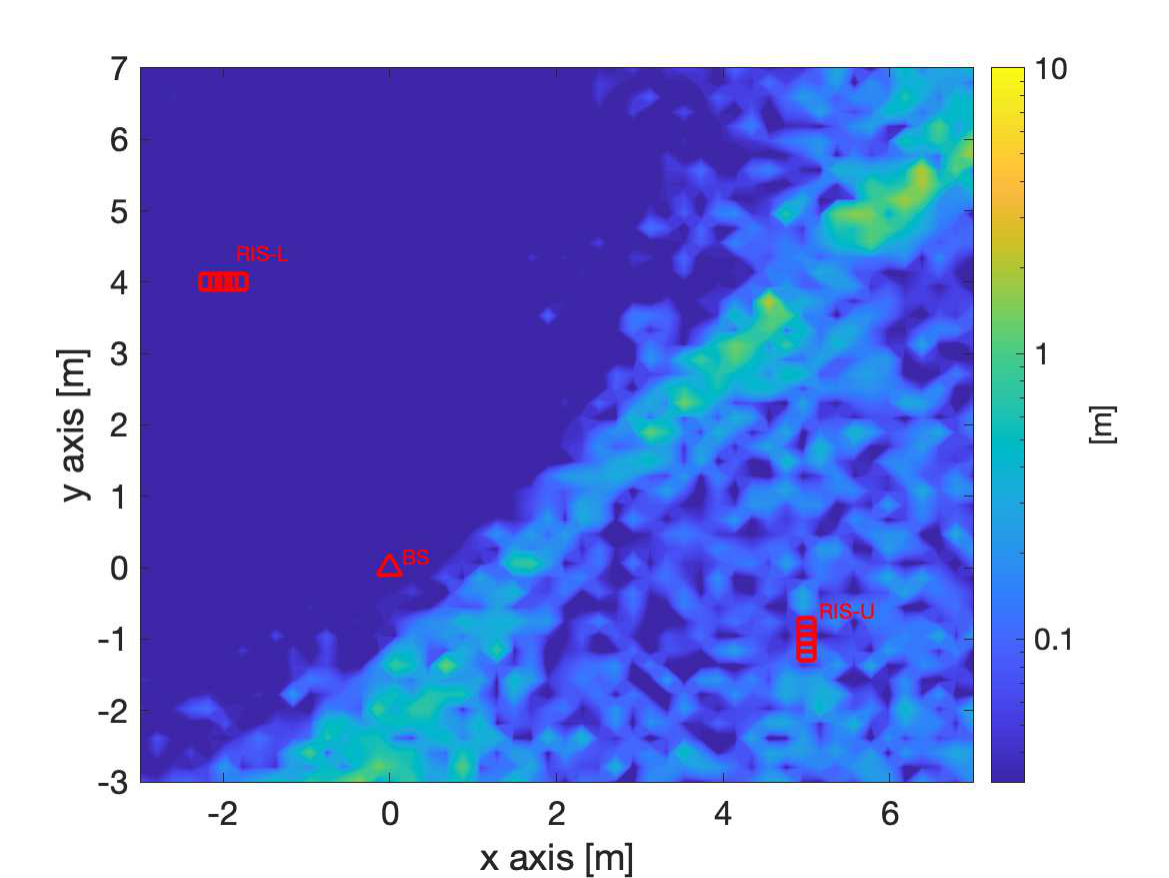} 
        \par Scenario 1 - RADC
    \end{minipage}
    \hfill
    \begin{minipage}[t]{0.32\textwidth}
        \centering
        \includegraphics[width=\textwidth]{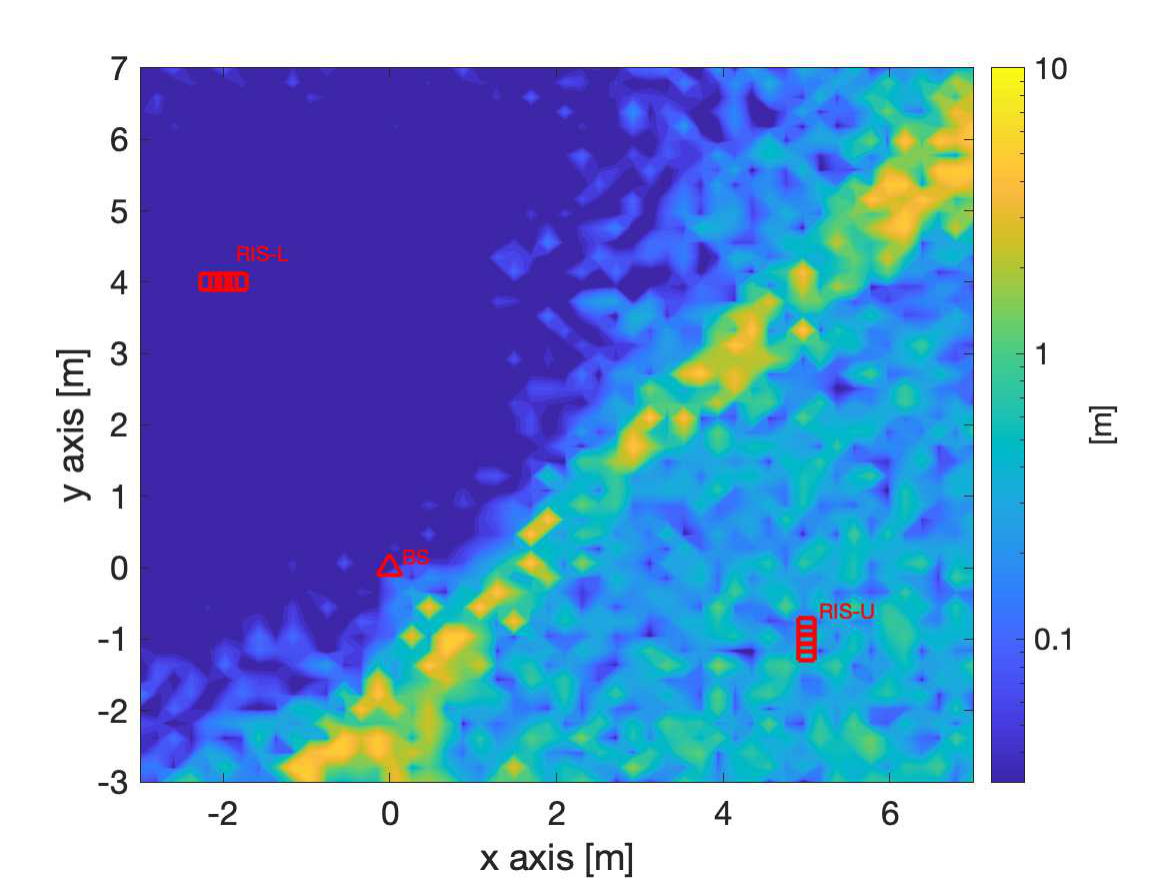} 
        \par Scenario 1 - RPDC
    \end{minipage}
    \vspace{0.5cm}
    \begin{minipage}[t]{0.32\textwidth}
        \centering
        \includegraphics[width=\textwidth]{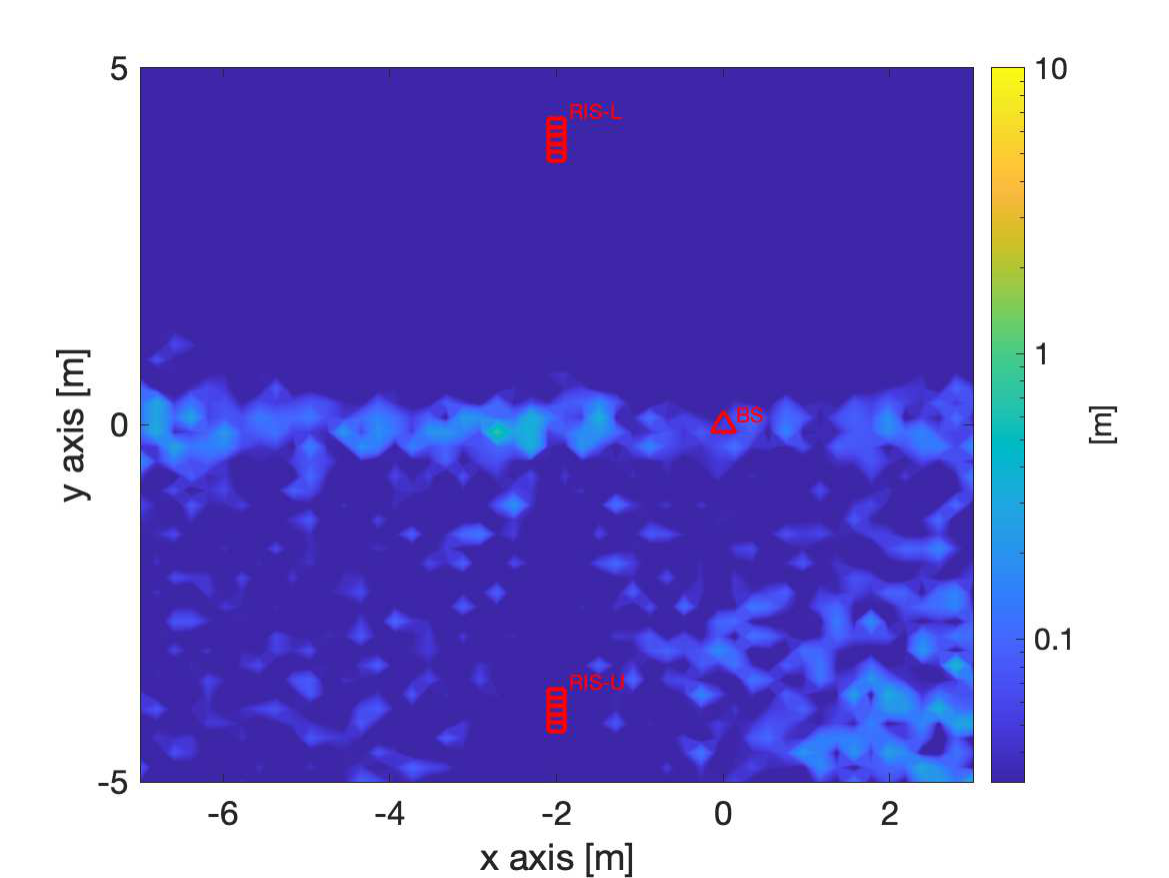} 
        \par Scenario 2 - random
    \end{minipage}
    \hfill
    \begin{minipage}[t]{0.32\textwidth}
        \centering
        \includegraphics[width=\textwidth]{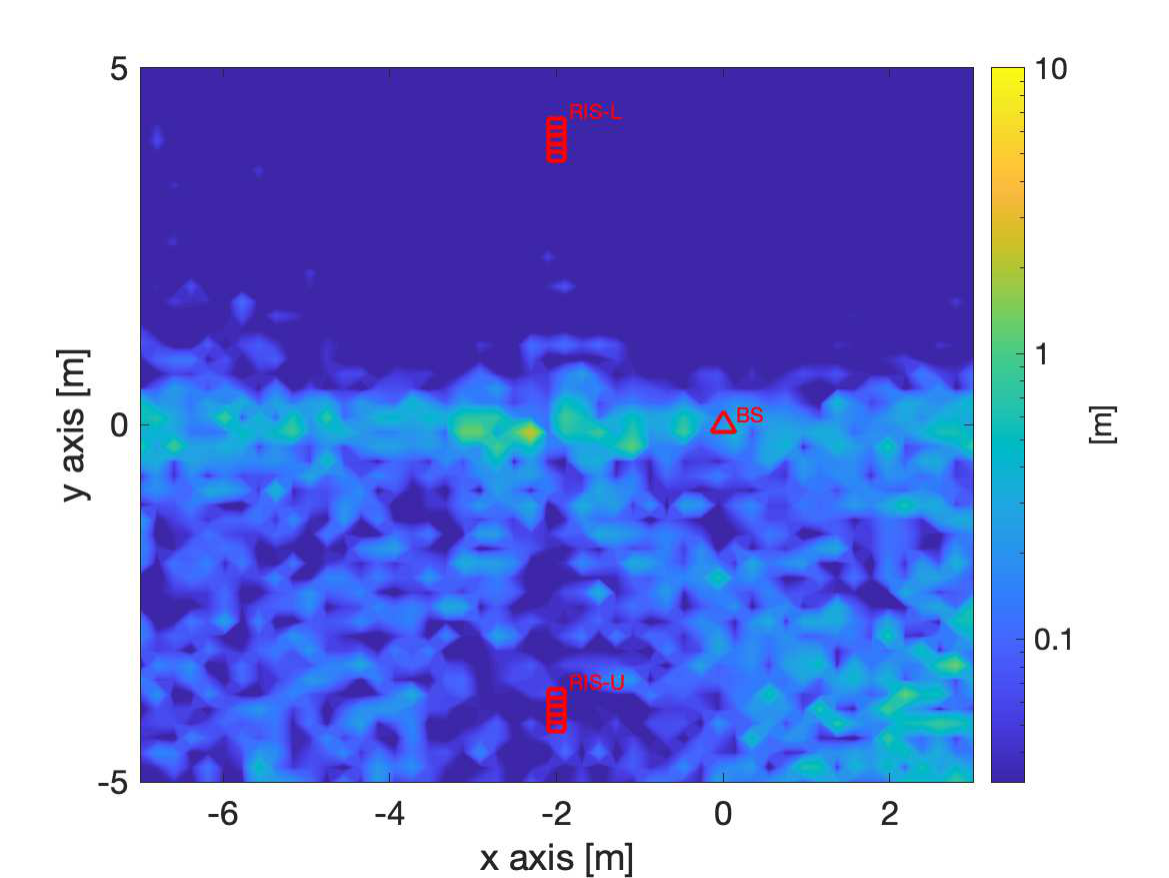} 
        \par Scenario 2 - RADC
    \end{minipage}
    \hfill
    \begin{minipage}[t]{0.32\textwidth}
        \centering
        \includegraphics[width=\textwidth]{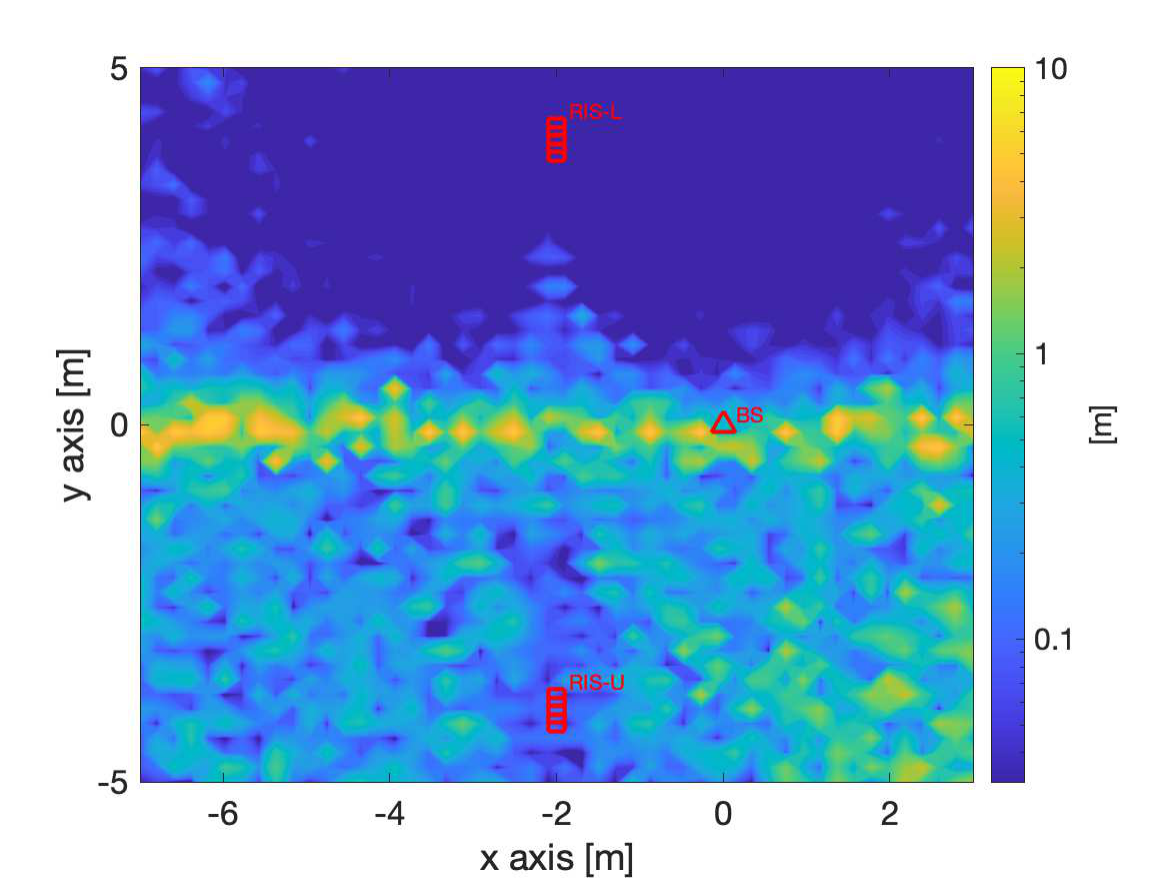} 
        \par Scenario 2 - RPDC
    \end{minipage}
    \vspace{0.5cm}
    \begin{minipage}[t]{0.32\textwidth}
        \centering
        \includegraphics[width=\textwidth]{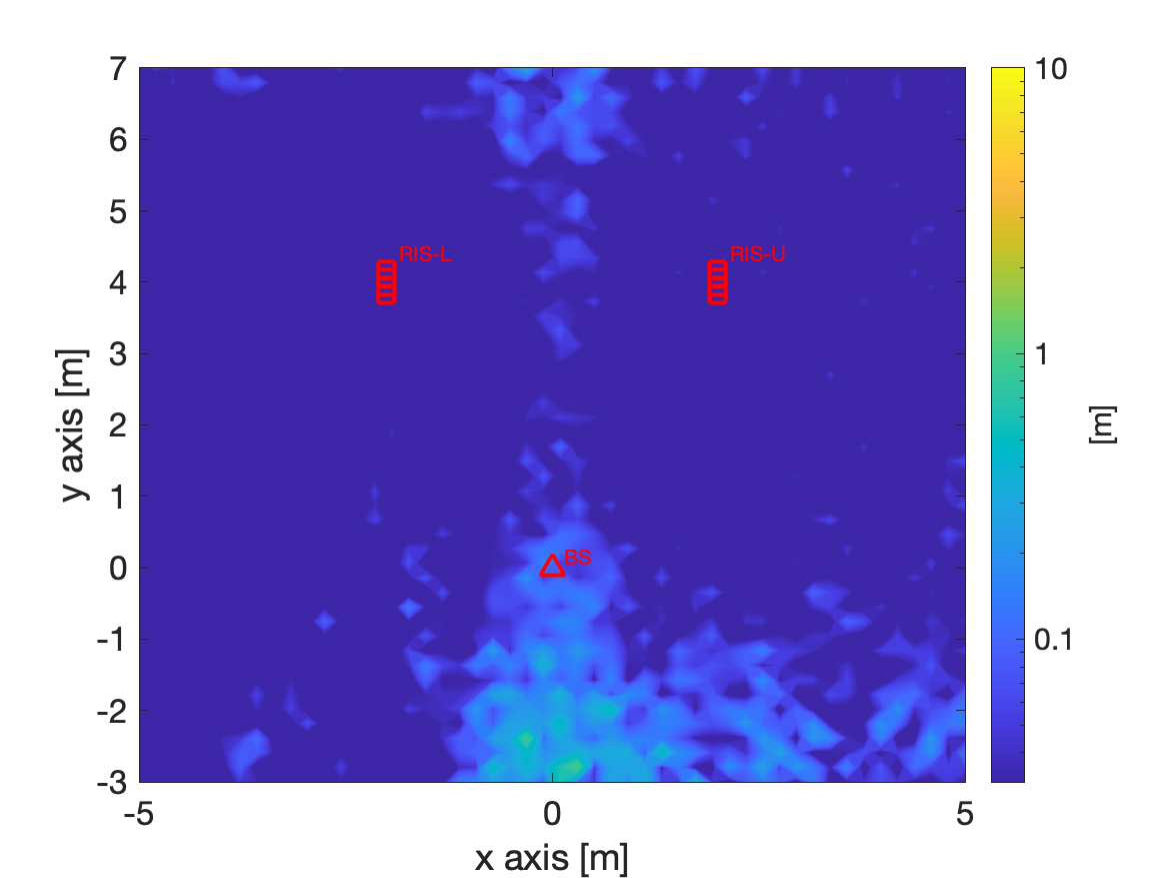} 
        \par Scenario 3 - random
    \end{minipage}
    \hfill
    \begin{minipage}[t]{0.32\textwidth}
        \centering
        \includegraphics[width=\textwidth]{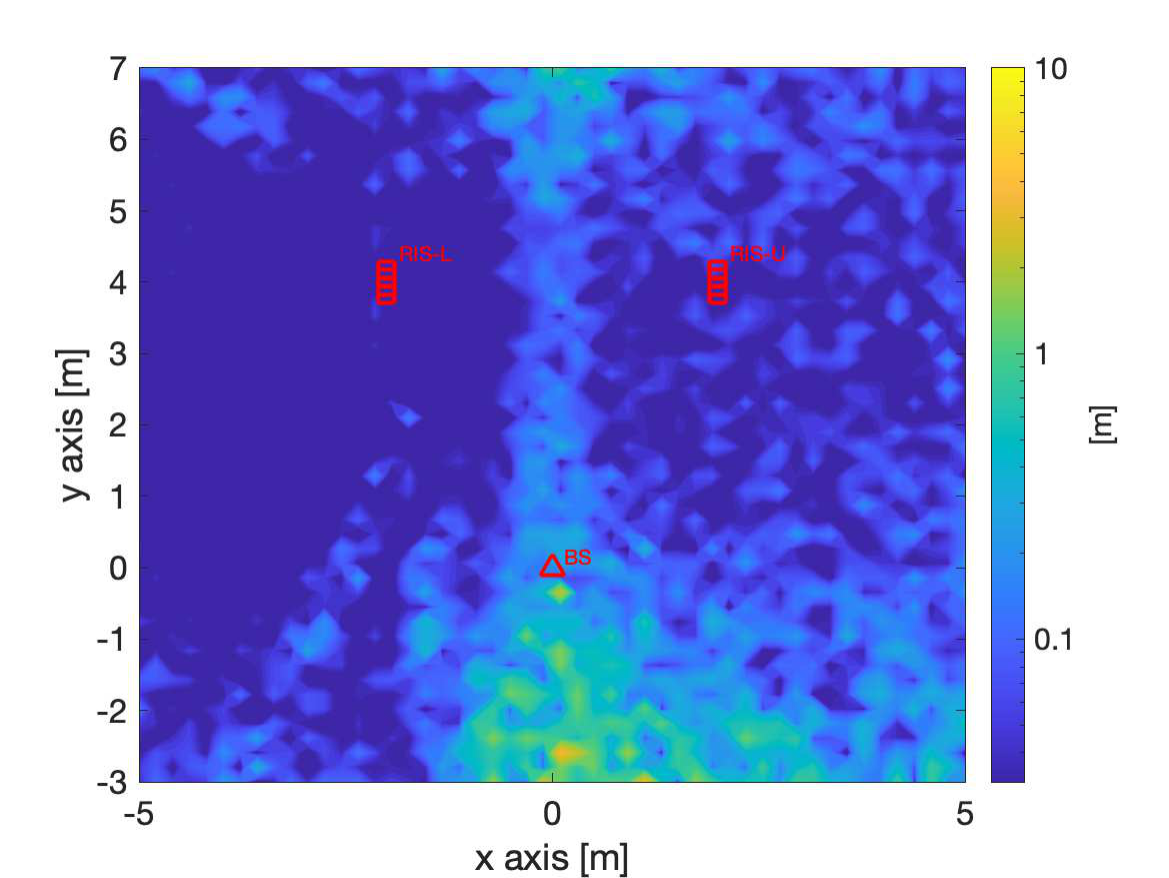} 
        \par Scenario 3 - RADC
    \end{minipage}
    \hfill
    \begin{minipage}[t]{0.32\textwidth}
        \centering
        \includegraphics[width=\textwidth]{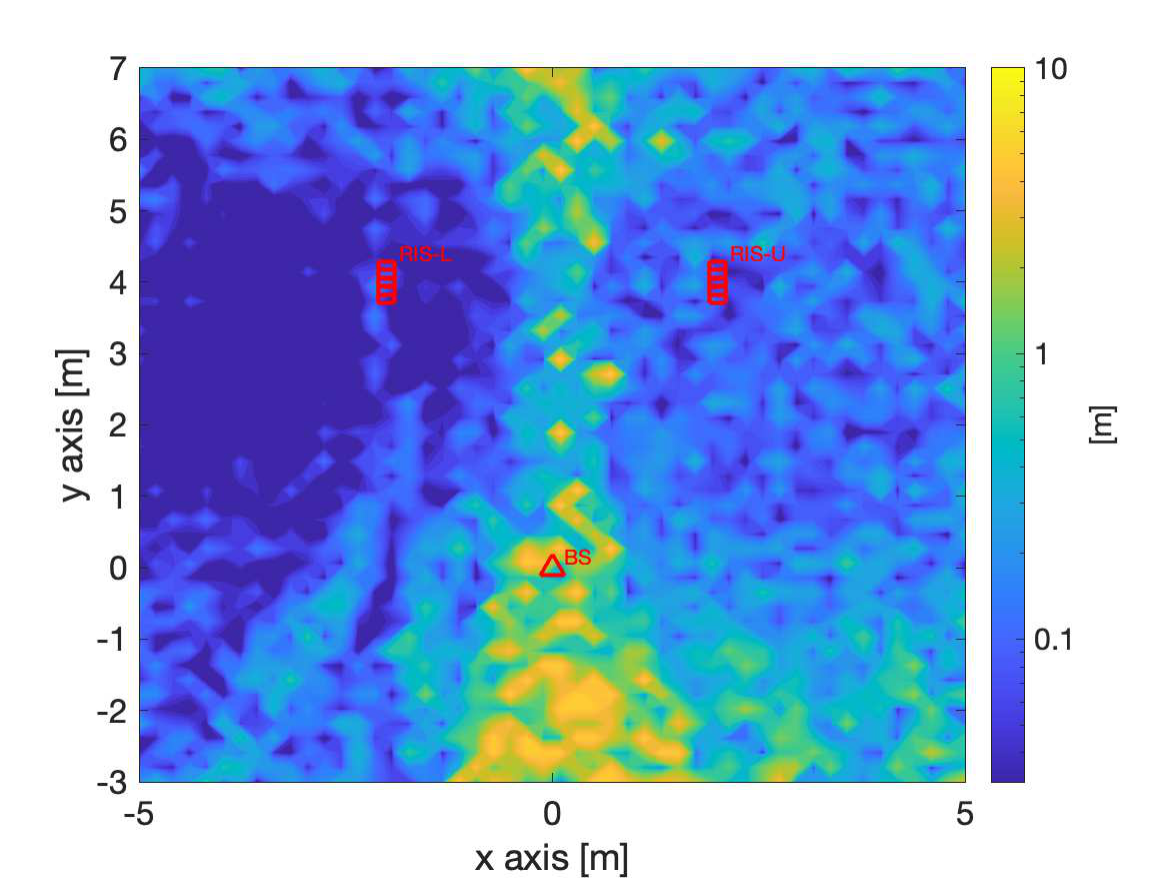} 
        \par Scenario 3 - RPDC
    \end{minipage}
   \caption{Visualization of $\text{ALB}_{\rm{pos}}$ for different UE locations within a $10 \times 10\ \text{m}^2$ area, with the BS fixed at $[0,0,0]^{\top}$. Two RISs are positioned on the same plane as the BS, 2 m above the UE. Three scenarios are considered, with results shown in rows 1, 2, and 3, respectively: Scenario 1 involves two RISs perpendicular to each other; Scenario 2 involves two RISs aligned along the same line facing the same direction; and Scenario 3 involves two RISs aligned along the same line but facing each other. Columns 1, 2, and 3 represent the results when the unauthorized RIS employs a random codebook, RADC, and RPDC, respectively.}
    \label{fig:ALB_UE}
\end{figure*}
\begin{figure}
    \centering
    \includegraphics[width=0.95\linewidth]{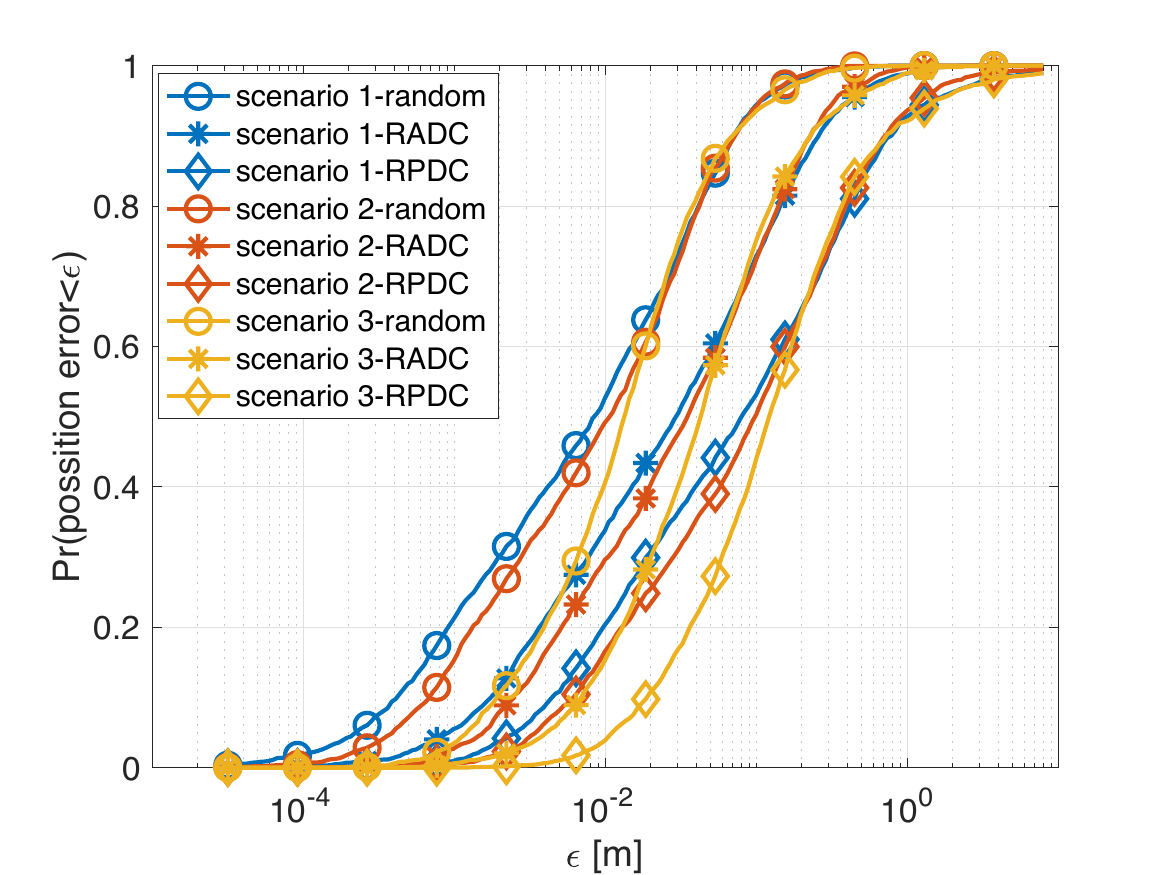}
    \caption{The CDF of the $\text{ALB}_{\rm{pos}}$ for three different scenarios.}
    \label{fig:CDF}
\end{figure}
\begin{figure}
    \centering
    \begin{minipage}[t]{0.46\textwidth} 
        \centering
        \includegraphics[width=\textwidth]{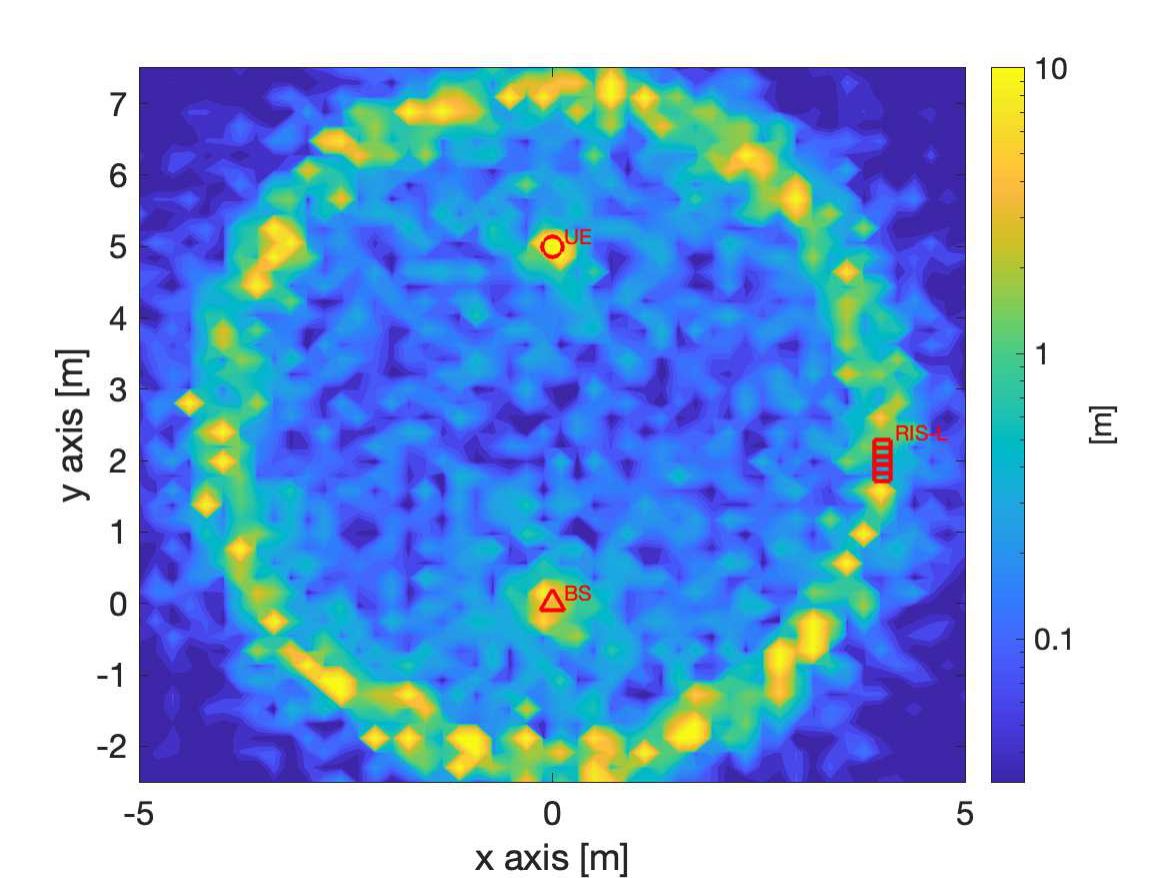} {\small{
        \*{(a)}}}
    \end{minipage}
    \hfill
    \begin{minipage}[t]{0.46\textwidth}
        \centering
        \includegraphics[width=\textwidth]{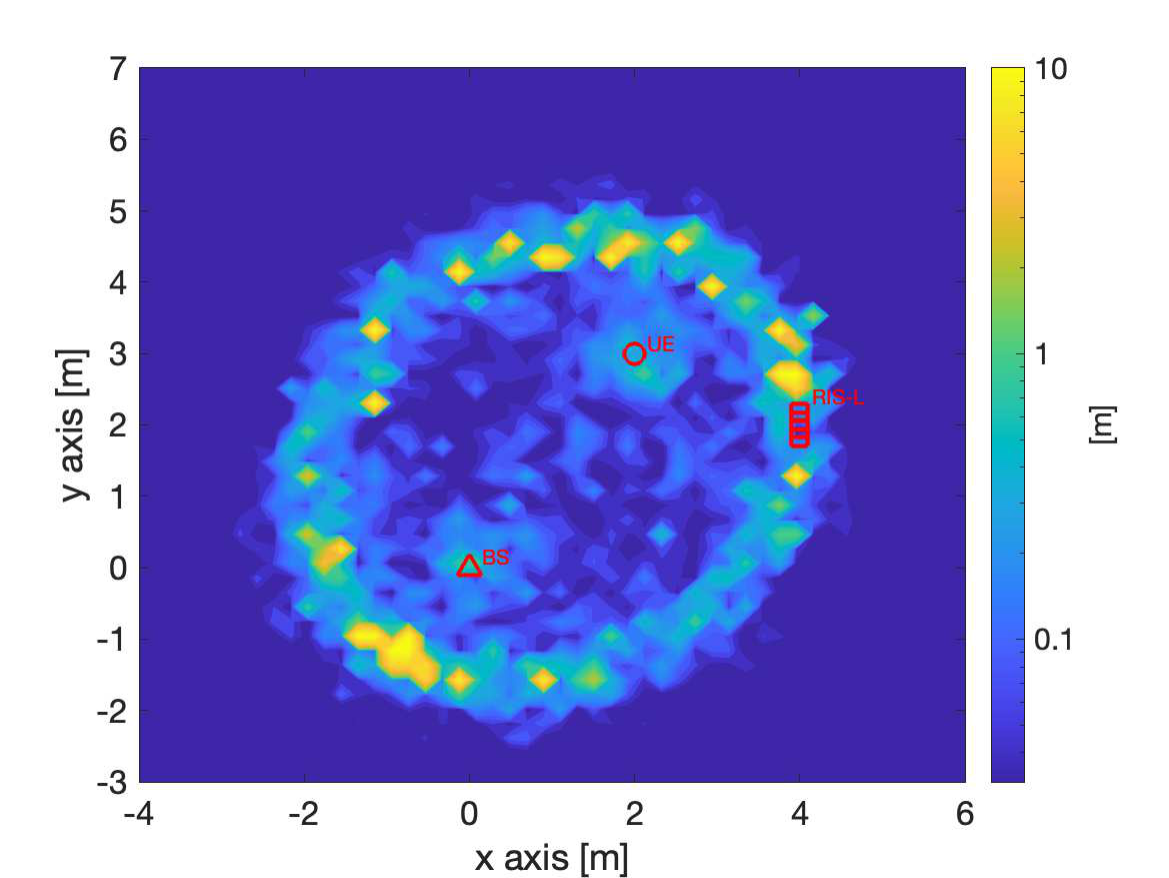} 
        {\small{
        \*{(b)}}}
    \end{minipage}
    \begin{minipage}[t]{0.46\textwidth}
        \centering
        \includegraphics[width=\textwidth]{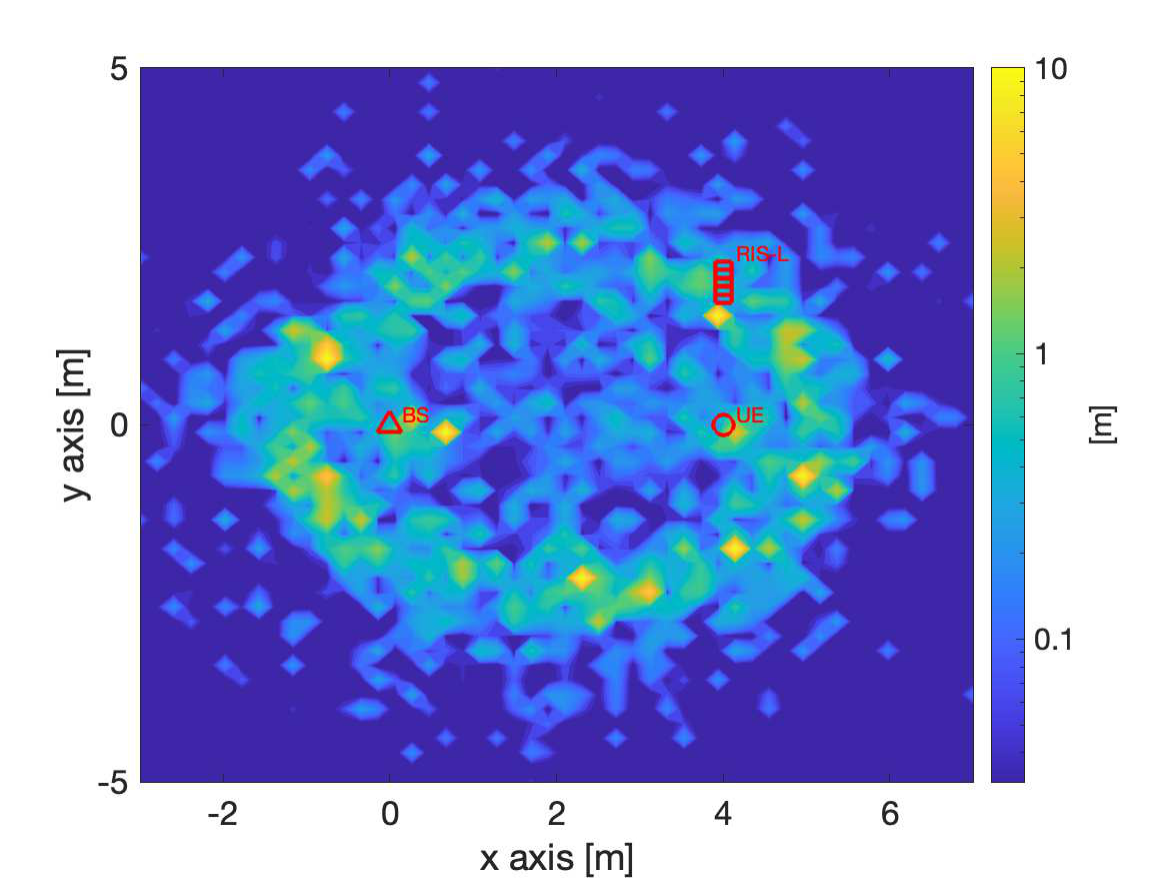} 
        {\small{
        \*{(c)}}}
    \end{minipage}
    \caption{Visualization of $\text{ALB}_{\rm{pos}}$ for different locations of an unauthorized RIS within a $10 \times 10 , \text{m}^2$ area, with the BS and legitimate RIS fixed at $[0,0,0]^{\top}$ and $[4,2,0]^{\top}$, respectively. Three different UE locations are considered: (a) UE and BS are vertically aligned; (b) BS and UE are aligned along a hypotenuse; (c) UE and BS are horizontally aligned. The unauthorized RIS employs the RPDC.}
    \label{fig:ALB_attacker}
\end{figure}
\subsection{The ALB with Different UE Locations }\label{subsec:ALB_UE}
With fixed BS and RIS locations, the positioning capability of an RIS-aided localization system varies as the UE location changes. Considering the presence of an unauthorized RIS, different relative positions of the legitimate and unauthorized RISs can result in different system scenarios. As discussed in Section \ref{subsec:ALB}, the ALB of the UE position can serve as a benchmark to evaluate the positioning capacity. From the point of view of assessing the performance of the localization system, we calculate the ALB for various UE locations within a $10 \times 10\ \text{m}^2$ area across three scenarios with different arrangements of the two RISs. While the legitimated RIS uses a random codebook, the results for unauthorized RIS using three different beamforming codebooks are investigated. 
The BS is fixed at $[0,0,0]^{\top}$ and the UE is placed in an unknown position in the $z = -2$ plane. The specific locations of the legitimate RIS and unauthorized RIS for the three scenarios are: (a) Scenario 1: Two RISs are perpendicular to each other with $\pv_{\rm {R_L}} = [-2, 4, 0]^\top$, $\pv_{\rm {R_U}} = [-2, 4, 0]^\top$; (b) Scenario 2: Two RISs are aligned along the same line and face the same direction with $\pv_{\rm {R_L}} = [-2, 4, 0]^\top$, $\pv_{\rm {R_U}} = [-2, 4, 0]^\top$; (c) Scenario 3: Two RISs are aligned along the same line and face each other with $\pv_{\rm {R_L}} = [-2, 4, 0]^\top$, $\pv_{\rm {R_U}} = [-2, 4, 0]^\top$. 

The results are shown in Fig. \ref{fig:ALB_UE}. By carefully analyzing the light-colored areas (which denote a higher ALB and poorer positioning performance) across the three scenarios, it becomes evident that UE positions resulting in a shorter unauthorized RIS path and a longer legitimate RIS path tend to have higher ALB. This is because the unauthorized RIS path exhibits a higher power level compared to the legitimate path, leading to stronger interference. Moreover, an obvious highlighted strip area can be observed in each scenario (particularly noticeable in the results using directional codebooks). These highlighted strip areas are caused by the close link distances of the legitimate RIS and unauthorized RIS paths. The poor positioning capability in these areas is not due to stronger signals from the unauthorized RIS. Instead, it is primarily attributed to ambiguity in the channel parameter optimization process with the MLE, caused by similar path delays. Identical or nearly identical delays in the RIS path make it less likely to achieve a clear and sharp global optimum for the cost function. 

In addition, the angle-resolving capability of a RIS plate diminishes when the incident angle is parallel to its plane. UE locations that result in large incident angles (relative to the local coordinate system of the legitimate RIS) and similar RIS path delay to that of the unauthorized RIS path at the same time are more likely to exhibit the highest ALB within the calculated area. For the results with different beamforming codebooks employed by unauthorized RIS, the ALB with the random codebook is smaller than 1 m, whereas the ALB with both directional codebooks exceeds 1 m. The positioning error when using directional codebooks may also exceed the limit of \( \sqrt{2 \times 2 \times 2} \) m, as the ALB is determined by the MLE-based optimization process, which does not confine the search within a constrained area like coarse estimation.
 Furthermore, it is evident that RPDC achieves more effective interference compared to RADC. The corresponding CDF of the ALB results, shown in Fig. \ref{fig:ALB_UE}, is presented in Fig. \ref{fig:CDF}, offering an alternative perspective on the positioning performance. As observed, the CDF results for Scenarios 1 and 2 are similar. The curves for random codebooks rise faster than those for directional codebooks, indicating that the random codebook achieves smaller positioning errors more consistently. At low positioning errors (e.g. \( \epsilon < 10^{-2} \, \text{m} \)), the random codebook significantly outperforms the directional codebooks. For larger errors (e.g. \( \epsilon \approx 1 \, \text{m} \)), the curves converge; however, the performance of RPDC remains the worst. For Scenario 3, the curves for all beamforming codebooks rise more slowly in the low position error region compared to the other scenarios, while they still converge as the error approaches \(1 \,\text{m}\). These observations demonstrate that the strongest interference is caused by unauthorized RIS using RPDC. Furthermore, the results align with the channel parameter estimation findings shown in Fig.~\ref{fig:ris_profiles}.


\subsection{The ALB with Different Locations of Unauthorized RIS }
Given that the original RIS-aided localization system has fixed locations for the legitimate RIS, BS, and UE, an interesting aspect to explore is how to strategically position the unauthorized RIS to generate intense interference. In the following, we calculate the ALB for three different scenarios as the location of unauthorized RIS changes within a $10 \times 10\ \text{m}^2$ area. Note that RISs, BS, and UE are positioned at the same height to facilitate an easier analysis of the underlying principles. The BS and legitimate RIS are fixed at $[0,0,0]^{\top}$ and $[4,2,0]^{\top}$, respectively and the locations of UE for three scebarios are: (a) Scenario 1: UE and BS are vertically aligned with $\pv_{\rm {U}} = [0, 5, 0]^\top$; (b) Scenario 2: UE and BS are aligned along ahypotenuse with $\pv_{\rm {U}} = [2, 3, 0]^\top$; (c) Scenario 3: UE and BS are horizontally aligned with $\pv_{\rm {U}} = [4, 0, 0]^\top$. The visualization of the positioning error as the unauthorized RIS changes its location is presented in Fig. \ref{fig:ALB_attacker}. Only the results with the unauthorized RIS using RPDC are provided, as similar patterns with lower ALB values are observed when using the other two codebooks.

For both scenarios, a highlighted elliptical ring is clearly observed, indicating that more effective interference can be generated if the unauthorized RIS is positioned along this elliptical ring. To further investigate the relationship between the observed ellipse and the system configuration, a general rule can be established: the observed elliptical ring aligns with an ellipse where the BS and the UE serve as the focal points, and the link distance of the legitimate RIS path corresponds to the major axis length of the ellipse. As shown in Fig. \ref{fig:ALB_attacker}, the orientation and size of the ellipse change based on the UE's location and the corresponding link distance of the legitimate RIS path. This observation is reasonable when considering the results from Fig.\ref{fig:ALB_UE}. Positioning the unauthorized RIS near the aforementioned ellipse results in similar link distances for both the legitimate RIS path and the unauthorized RIS path. As discussed in Section~\ref{subsec:ALB_UE}, this similarity leads to poor positioning performance due to increased ambiguity in parameter estimation. Moreover, a higher ALB can also be observed when the unauthorized RIS is positioned close to either the BS or the UE, thereby introducing a high-power unauthorized RIS path. If the malicious agent aims to degrade the positioning performance of two or more UEs, the optimal location for the unauthorized RIS would be at the intersections or overlapping areas of the ellipses corresponding to the different UEs. The above discussions could also be valuable in preparing effective strategies to mitigate unexpected interference.

\section{Conclusion}  
This paper investigates RIS-aided positioning in the presence of interference from an unauthorized RIS. We formulate the positioning problem and propose two localization algorithms: a low-complexity estimator and the MLE. Theoretical analysis and numerical simulations are performed to assess the impact of unauthorized RIS interference on both channel estimation and positioning performance and validate the proposed algorithms.  The results indicate that, under interference, localization performance becomes saturated at a certain level as the SNR increases because of the model mismatch. Moreover, significant estimation errors occur when the unauthorized RIS path is strong or when its delay aligns closely with that of the legitimate RIS. It is also observed that an unauthorized RIS with greater system knowledge can induce higher estimation errors by employing directional beams, compared to using a random beamforming codebook.  Future work should focus on developing effective countermeasures for detecting and mitigating interference, as well as localizing malicious sources to enhance the robustness and reliability of RIS-aided positioning systems.

\bibliographystyle{IEEEtran}
\bibliography{ref}


 




\vfill

\end{document}